\documentclass[aps,prd,nofootinbib,twocolumn,superscriptaddress,preprintnumbers,letterpaper,natbib,nofootinbib]{revtex4-1}
\pdfoutput = 1
\usepackage{amssymb}
\usepackage{amsmath}
\usepackage{epsfig}
\usepackage{hyperref}
\usepackage{color}
\usepackage{cleveref}
\usepackage{multirow}
\usepackage{capt-of}
\usepackage{siunitx}
\usepackage{graphicx}
\usepackage[caption=false]{subfig}
\usepackage{slashed}
\usepackage{tabularx}
\usepackage{enumitem}
\usepackage{multirow}

\newcommand{\be}{\begin{equation}}
\newcommand{\ee}{\end{equation}}
\newcommand{\bea}{\begin{eqnarray}}
\newcommand{\eea}{\end{eqnarray}}

\newcommand{\prn}[1]{ \left(  #1 \right) }

\newcommand{\al}[1]{\begin{align} #1 \end{align}}

\newcommand{\Neff}{N_{\text{eff}}}
\newcommand{\gamp}{{\gamma'}}

\definecolor{sky}{rgb}{0., 0.5883488108126352, 0.9445016153846155}
\definecolor{cadmiumgreen}{rgb}{0.0, 0.42, 0.24}

\begin{document}

\title{Asymmetric Dark Matter May Not Be Light}

\author{Eleanor Hall}
\affiliation{Berkeley Center for Theoretical Physics, University of California, Berkeley, CA 94720, USA}
\affiliation{Theory Group, Lawrence Berkeley National Laboratory, Berkeley, CA 94720, USA}

\author{Robert McGehee}
\affiliation{Leinweber Center for Theoretical Physics, Department of Physics,
University of Michigan, Ann Arbor, MI 48109, USA}

\author{Hitoshi Murayama}
\affiliation{Berkeley Center for Theoretical Physics, University of California, Berkeley, CA 94720, USA}
\affiliation{Theory Group, Lawrence Berkeley National Laboratory, Berkeley, CA 94720, USA}
\affiliation{Kavli Institute for the Physics and Mathematics of the Universe (WPI), University of Tokyo, Kashiwa 277-8583, Japan}

\author{Bethany Suter}
\affiliation{Berkeley Center for Theoretical Physics, University of California, Berkeley, CA 94720, USA}
\affiliation{Theory Group, Lawrence Berkeley National Laboratory, Berkeley, CA 94720, USA}

\begin{abstract} 
It is often said that asymmetric dark matter is light compared to typical weakly interacting massive particles. Here we point out a simple scheme with a neutrino portal and $\mathcal{O}(60 \text{ GeV})$ asymmetric dark matter which may be ``added'' to any standard baryogenesis scenario. The dark sector contains a copy of the Standard Model gauge group, as well as (at least) one matter family, Higgs, and right-handed neutrino. After baryogenesis, some lepton asymmetry is transferred to the dark sector through the neutrino portal where dark sphalerons convert it into a dark baryon asymmetry. Dark hadrons form asymmetric dark matter and may be directly detected due to the vector portal. Surprisingly, even dark anti-neutrons may be directly detected if they have a sizeable electric dipole moment. The dark photons visibly decay at current and future experiments which probe complementary parameter space to dark matter direct detection searches. Exotic Higgs decays are excellent signals at future $e^+ e^-$ Higgs factories.
\end{abstract}
\maketitle

\section{Introduction}
\label{sec:Intro}
The dark matter (DM) enigma has come to captivate experts in astrophysics, cosmology, and particle physics. Despite determined efforts to detect its interactions with Standard Model (SM) particles, both directly~\cite{Aprile:2018dbl} and indirectly~\cite{Leane:2018kjk}, little is known about it except that it's abundant and not part of the SM.

The baryon asymmetry of the Universe (BAU) is another cosmic mystery which requires an explanation involving physics beyond the Standard Model (BSM)~\cite{Jarlskog:1985ht,Gavela:1994dt,Huet:1994jb,Kajantie:1995kf,Kajantie:1996qd,Rummukainen:1998as}. Many solutions have been posited, though they typically fall within one of two broad classes: \emph{leptogenesis}~\cite{Fukugita:1986hr} or \emph{electroweak baryogenesis}~\cite{Cohen:1993nk,Morrissey:2012db,Konstandin:2013caa}.\footnote{There also exist more exotic ideas, such as \emph{mesogenesis}~\cite{Elor:2018twp,Alonso-Alvarez:2019fym,Elor:2020tkc}, but these are incompatible with the DM model presented here.} Models of the former are quite minimal if one is already inclined to invoke the seesaw mechanism to explain neutrino masses~\cite{Minkowski:1977sc,GellMann:1980vs,Yanagida:1980xy}, but are difficult to probe experimentally due to the high scale of new physics (NP) -- often $10^{10} \text{ GeV}$ or higher~\cite{Buchmuller:2004nz}. Arguably the best probe would be gravitational wave from cosmic strings associated with the phase transition that generates right-handed neutrino mass \cite{Dror:2019syi}. Models of the latter are often too experimentally probe-able, with many ruled out by null results at the Large Hadron Collider as well as increasingly precise electric dipole moment (EDM) measurements~\cite{Andreev:2018ayy}.

Nonetheless, there still exist several promising iterations of electroweak baryogenesis with viable parameter space that may be probed in the near future. For example, models of split supersymmetry, 2 Higgs doublets, real scalar singlet extensions, and composite Higgs all provide viable explanations of the BAU originating from the electroweak phase transition (see \emph{e.g.}~\cite{Demidov:2016wcv,Dorsch:2016nrg,Vaskonen:2016yiu,Bruggisser:2018mrt} for recent work in each direction).

In this paper, we propose a dark sector which may be added onto any of these existing solutions to generate asymmetric dark matter (ADM) without impacting the BAU production. It is general and does not care about the specifics of the BAU-setting mechanism; we only assume the initial SM baryon and lepton asymmetries are equal, as is true in electroweak baryogenesis realizations.

In the minimal scenario, the dark sector consists of a dark $SU(3)' \times SU(2)' \times U(1)'$ gauge group, a dark generation of matter (including a right-handed Weyl neutrino), and a dark Higgs doublet, though there may be more of the latter two. This simple dark sector realizes ADM in the following way. After the SM electroweak phase transition (EWPT), equal SM baryon and lepton asymmetries are generated, for example, by one of the scenarios listed above. The neutrino portal allows some of the SM lepton asymmetry to transfer to a dark lepton asymmetry. Dark sphalerons from the dark $SU(2)$ anomaly then convert some of this asymmetry into a dark baryon asymmetry, which is conserved after the dark EWPT. The resulting asymmetric dark hadrons form DM.\footnote{For the ``opposite'' possibility in which a dark EWPT causes a dark lepton asymmetry to transfer to the SM where it is converted into SM baryons via sphalerons, see \emph{e.g.} \cite{Hall:2019ank,Hall:2019rld}. In this case, the ADM is light $\sim O(1)$~GeV.}

To shed excess entropy from the symmetric component of dark hadrons before big bang nucleosynthesis (BBN), we use the kinetic mixing between $U(1)'$ and $U(1)_Y$. This vector portal causes the dark photon to decay to the SM bath prior to SM $\nu$ decoupling near $3 \text{ MeV}$~\cite{Mangano:2006ar}. Since the dark photons only visibly decay, there are a number of current and proposed searches which may detect them. The dark hadrons, even dark anti-neutrons with a sizable EDM, also directly scatter off protons via dark photons in direct detection experiments. 

The organization of this paper is as follows. In Sec.~\ref{sec:darksec}, we detail the dark sector as well as the minimal assumptions of the mechanism responsible for baryogenesis. We then calculate the transfer and conversion of SM asymmetries into dark sector asymmetries in Sec.~\ref{sec:transfer}. Signals of this scenario due to the visibly decaying dark photon and direct detection are discussed in Sec.~\ref{sec:signals}. We conclude with a discussion of future directions in Sec.~\ref{sec:discuss}.

\section{The Dark Sector}
\label{sec:darksec}
The minimal dark sector mimics the SM and consists of a $S U(3)^{\prime} \times S U(2)^{\prime} \times U(1)^{\prime}$ gauge group, one 
matter generation (with a right-handed Weyl neutrino), and one Higgs doublet, which we denote:
\begin{equation}
    Q^{\prime}, u_{\mathrm{R}}^{\prime}, d_{\mathrm{R}}^{\prime}, L^{\prime}, e_{\mathrm{R}}^{\prime}, N_{\mathrm{R}}, H'. \nonumber
\end{equation}
Throughout, superscripts $'$ are attached to the dark sector equivalents of SM particles and couplings. For generality, we allow the dark Yukawa couplings to have different hierarchies than in the SM and examine several different limiting cases. 

There exist two portals between the dark and SM sectors. The vector portal comes from the kinetic mixing of dark and SM hypercharge, which after EWPTs in both sectors results in 
\begin{equation}
    \mathcal{L} \supset \frac{\epsilon}{2} F_{\mu \nu} F^{\prime \mu \nu}+\frac{1}{2} m_{\gamma^{\prime}}^{2} A_{\mu}^{\prime} A^{\prime \mu}.
\end{equation}
Note that we additionally include a mass term for the dark photon to permit its necessary decay to SM fermions. The neutrino portal is provided by the right handed neutrino's couplings to both dark and SM Higgses:
\begin{equation}
    \mathcal{L} \supset y'_{N} \bar{L}^{\prime} \tilde{H}' N_{\mathrm{R}} + y_{N} \bar{L} \tilde{H} N_{\mathrm{R}}+\text { c.c. }.
\end{equation}
Here and below, we take $N_{\mathrm{R}}$ to be a Weyl fermion which is the minimal choice yet different from a Dirac fermion adopted by the previous analyses \cite{Hall:2019ank,Hall:2019rld}.
We note that there could be a Majorana mass for the dark neutrino, but we assume it is small. We also assume that the neutrino portal transfer rate is slow enough such that no asymmetry leaks over into the dark sector lepton number until after the SM phase transition is complete, which is possible by taking $y_N$ sufficiently small.

As mentioned above, we assume there is a mechanism of baryogenesis and for simplicity, that it creates equal SM lepton and baryon asymmetries, $L$ and $B$. We discuss a more general case without this assumption in Appendix \ref{sec:lepbarasym}. One might also consider adding this dark sector onto a model of leptogenesis. However, there appears to be no motivated reason why any such model would not also allow an arbitrary dark lepton asymmetry to be generated at the same time as the SM lepton asymmetry. Because of this freedom, the asymmetries of the dark and SM sectors would no longer be related and the DM mass scale would no longer be predicted. We therefore do not consider this possibility further.

\begin{figure*}[t!]
\includegraphics[width =\columnwidth]{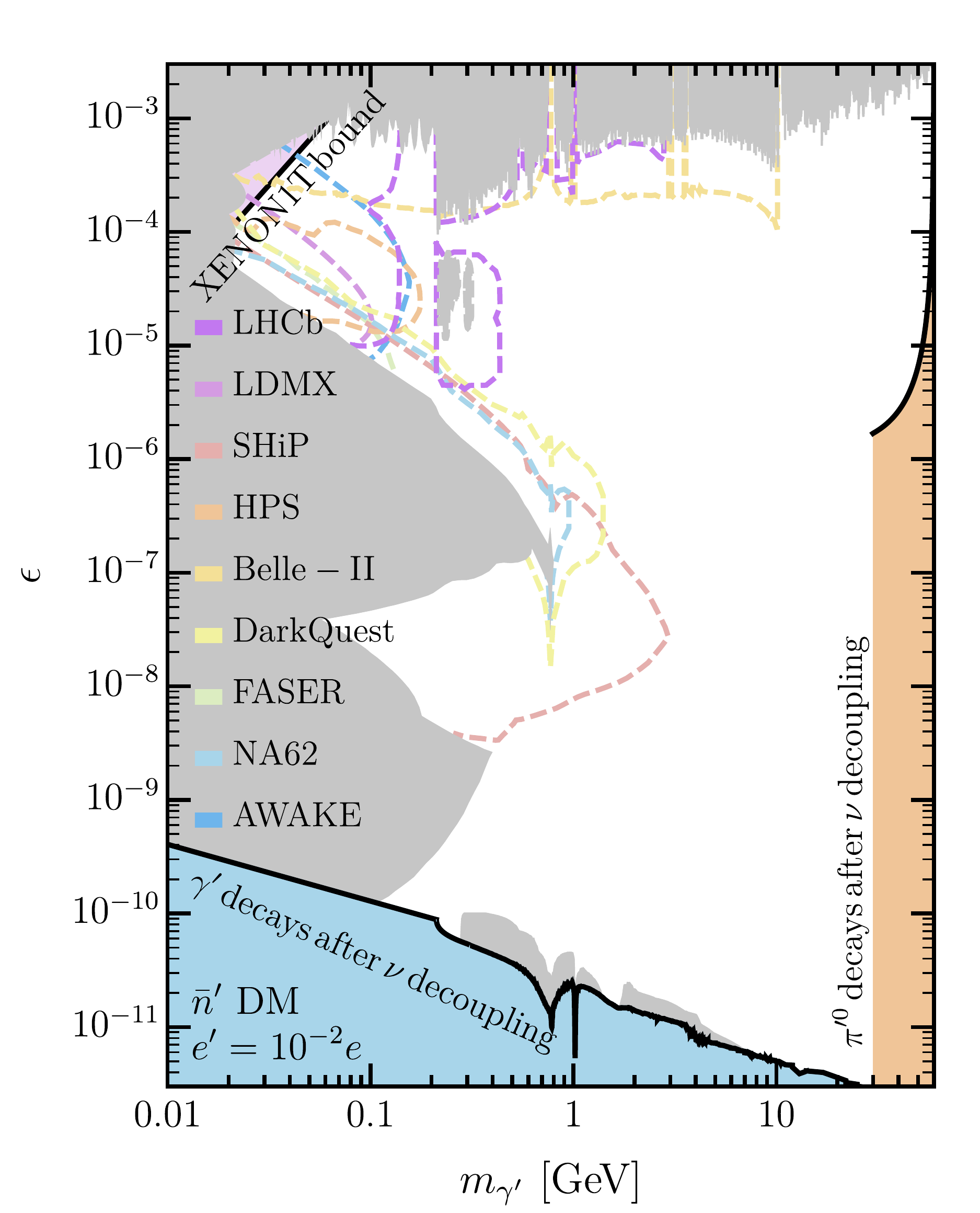}
\includegraphics[width =\columnwidth]{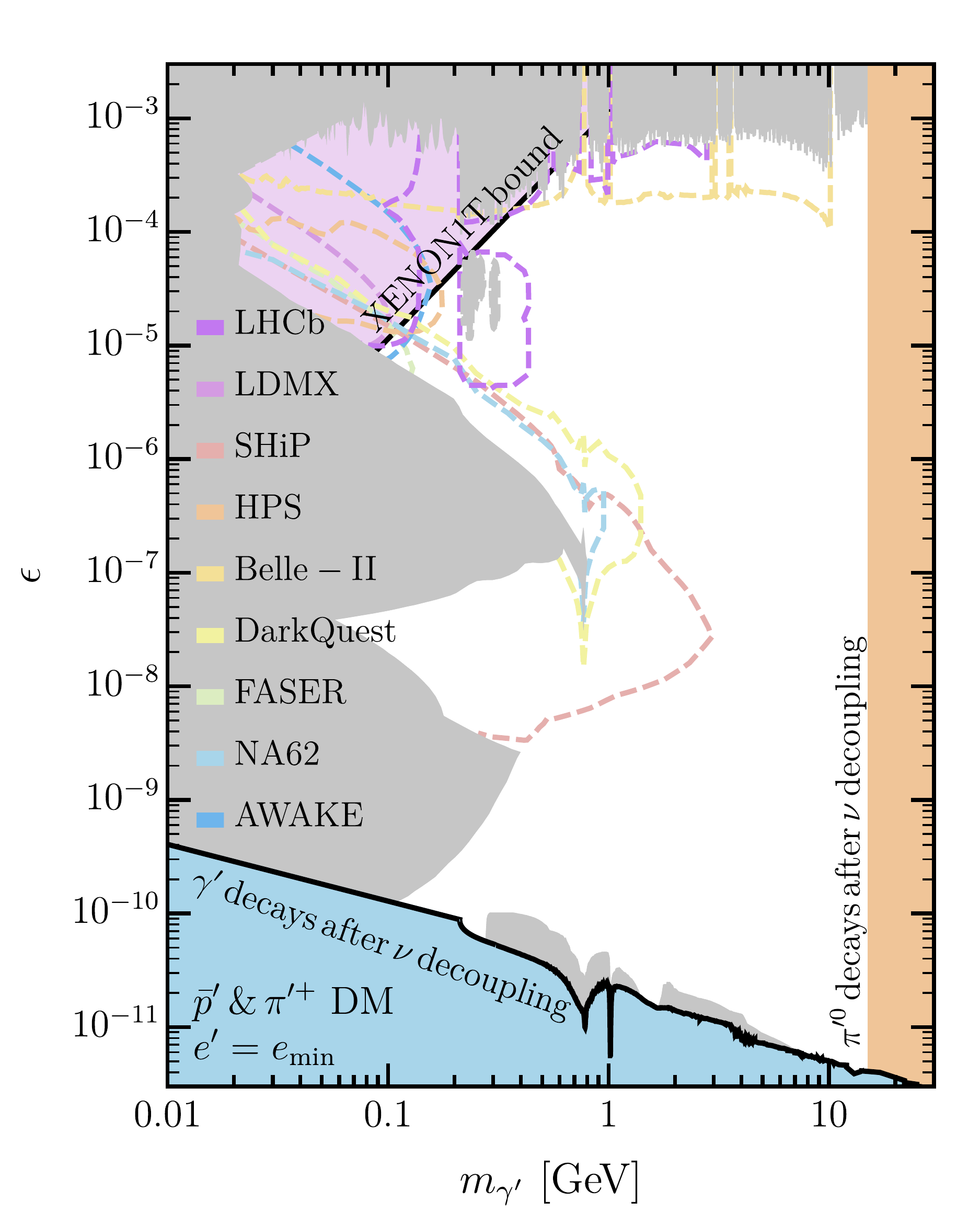}
\caption{\label{fig:darkphoton} The viable ADM parameter space as projected onto dark photon mass versus kinetic mixing. Existing constraints on visibly decaying dark photons~\cite{Fradette:2014sza,Alexander:2016aln,Chang:2016ntp,Hardy:2016kme,Pospelov:2017kep,Banerjee:2018vgk,Aaij:2017rft,Aaij:2019bvg,Parker:2018vye,Tsai:2019mtm} are shaded dark gray, while projected sensitivities are dashed~\cite{Celentano:2014wya,Ilten:2015hya,Alekhin:2015byh,Ilten:2016tkc,Alexander:2016aln,Caldwell:2018atq,Kou:2018nap,Berlin:2018pwi,Berlin:2018bsc,Ariga:2018uku,NA62:2312430,Tsai:2019mtm}. Color-shaded regions are ruled out by too-late decays and direct detection constraints~\cite{Aprile:2018dbl}, as discussed in the text. {\bf Left:} DM is all $\bar{n}'$ and $e'=10^{-2} e$. {\bf Right:} DM is 50\% $\bar{p}'$ and 50\% $\pi'^+$ and $e'=e_{\text{min}}$.} 
\end{figure*}

\section{Asymmetry Transfer}
\label{sec:transfer}

Since we have assumed that there is a SM electroweak baryogenesis scenario which generates $B$ and $L$, we must now describe how some of that asymmetry is ported over to the dark sector to form ADM. In all but the specific case of extremely massive charged dark leptons discussed later, the neutrino portal remains active until after the dark EWPT. We use the $N_R - H$ Yukawa to convert a SM neutrino (with $L = 1$) into a $N_R$ (with $L' = 1$). Although the Higgs mediator is off shell below the EWPT, it should still be able to scatter off any SM fermions in the bath. Since the dark EWPT occurs at a temperature above $\Lambda'_{QCD}$ and $\Lambda'_{QCD} \sim m_\chi \gtrsim 5$ GeV, all SM leptons are in the bath prior to the dark EWPT. Therefore, the two most relevant processes for $L \leftrightarrow L'$ after the EWPT are $\nu_i + e_i \leftrightarrow N_R + e_i$ and $e_i^- + u_j \leftrightarrow N_R + d_j$. We also note that the dark right-handed neutrino will interact with the dark sector Higgs via $h'^0 \leftrightarrow N_R + \nu_i'$.

Using standard chemical equilibrium calculations~\cite{Harvey:1990qw}, the baryon numbers, lepton numbers, and charges may be written in terms of chemical potentials after the SM EWPT as
\begin{align}
\label{eq:BLQ}
     B & =  6 (\mu_u + \mu_d), \nonumber \\
     B' & =  4 \mu_{uL'} + 2 \mu_{W'}, \nonumber \\
     L & =  9 \mu_{N_R} + 6 (\mu_d - \mu_u), \\
     L' & =  3\mu' + 2 \mu_{W'} - \mu_{0'} + \mu_{N_R}, \nonumber \\
     Q & =  24 \mu_u - 18 \mu_d - 6 \mu_{N_R}, \nonumber \\
     Q' & =  2 \mu_{uL'} - 2 \mu' - 10 \mu_{W'} + 6 \mu_{0'}, \nonumber
\end{align}  
where $\mu'$ and $\mu_{0'}$ are the chemical potentials of the left-handed dark neutrino and neutral dark Higgs, respectively. In deriving the above for the most minimal dark sector, we have set the number of dark sector families, Higgs, and right-handed neutrinos to one. The more general case is considered in Appendix~\ref{sec:lepbarasym}. Additionally, while sphalerons and the neutrino portal in the dark sector are active, they impose
\al{
\label{eq:sphal}
&3 \mu_{uL'}+2\mu_{W'}+\mu'=0 & &
\mu_{0'}=\mu'-\mu_{N_R}.
}

Assuming that the neutrino portal is active until after the dark EWPT and that it is strongly first order, we find
\begin{equation}
    B' = -\frac{72}{535}B, \hspace{5 mm} L' = \frac{168}{535}B
\end{equation}
If we instead assume that the dark EWPT is a crossover, we find
\begin{equation}
\label{eq:dPT2ndOrd}
    B' = -\frac{120}{1427}B, \hspace{5 mm} L' = \frac{360}{1427}B.
\end{equation}
Either scenario is viable and only differs from the other slightly in the predicted DM mass.

The calculation above assumes that the W bosons are still approximately massless at the time of the dark EWPT. This is not strictly necessary for some lighter DM masses we consider below which may come from dark sectors with lower EWPT temperatures. However, we find that in the case that the SM W's have already left the bath at the time of the dark EWPT, the asymmetries only differ by less than $5\%$.

As mentioned above, we can also consider the case where we give the right-handed neutrino a finite mass. As we increase its mass relative to the temperature of the dark EWPT, the ability of the neutrino portal to transfer the SM lepton asymmetry to a dark one diminishes. Additionally, the dark sphaleron rate will slowly turn off, again, acting to decrease the number density of dark baryons. Thus, as the mass of the right-handed neutrino increases, the mass of DM also increases to compensate for the smaller dark baryon asymmetry. At some point not too far above the dark EWPT, the neutrino mass will be so heavy that it necessitates a DM mass above the SM EWPT temperature, corresponding to a breakdown of the assumptions of this scenario.

Future work is needed for a more quantitative analysis. The main difficulty lies in calculating the modifications to the dark sphaleron rate as a function of right-handed neutrino mass. Since calculations of the sphaleron rate in the SM are intrinsically related to its own parameters~\cite{DOnofrio:2012phz, DOnofrio:2014rug, Rubakov:1997sr}, it is difficult to immediately generalize any known results to this dark sector. The freedom of many parameters in the dark sector also complicates any possible quantitative calculation.

\section{Visible Signals}
\label{sec:signals}

Having established the minimal content of the dark sector and the predicted dark baryon asymmetry, we now outline the two main possible DM scenarios and their corresponding visible signals. For the minimal case of one dark sector matter generation, we have the freedom to choose the masses of the two light quarks. If $m_{d'}<m_{u'}$, then the mass of the dark anti-neutron, $m_{\bar{n}'}$, will be lighter than that of the dark anti-proton, $m_{\bar{p}'}$. Consequently, all of the dark baryon asymmetry will be held by dark anti-neutrons which will therefore comprise the entirety of DM. In the simplest case of a crossover dark EWPT, given by Eq.~\eqref{eq:dPT2ndOrd}, we find:
\al{
\label{eq:mnp}
m_{\bar{n}'} = 59.9 \text{ GeV}.
}

Alternatively, if $m_{u'}<m_{d'}$, then the dark anti-proton is the lightest dark baryon and DM is comprised of equal numbers of dark anti-protons and pions. To permit the greatest range of dark photon masses, we consider the limit in which the dark pion mass approaches the dark anti-proton mass. Thus, again for the case of a crossover dark EWPT, we have
\al{
\label{eq:mpp}
m_{\bar{p}'}=m_{\pi'}=29.9 \text{ GeV}.
}

Interestingly, this ADM in both scenarios is heavy relative to what one naively suspects. For $\mathcal{O}(1)$ equal asymmetries between the dark and SM sectors, one would expect an $\mathcal{O}(5 \text{ GeV})$ DM. In practice, many models have slightly larger asymmetries in the dark side, pushing ADM masses more towards $\mathcal{O}(1 \text{ GeV})$. To our knowledge, this model of ADM is thus heavier than usual.

In both of these DM scenarios, the DM may directly scatter off charged SM matter, such as the protons in Xenon nuclei within direct detection experiments. The scattering cross section for the $\bar{p}'-\pi'^{+}$ DM case is
\al{
\sigma_{\chi p} \approx \epsilon^2 e^2 e^{\prime 2} 
\frac{m_p^2 m_{\bar{p}'}^2}{\pi (m_p+m_{\bar{p}'})^2 m_{\gamma'}^4},
}
where $\chi$ here refers to either DM sub-component, $\bar{p}'$ or $\pi'^{+}$.
For the case of $\bar{n}'$ DM, it may scatter if dark QCD has a sizable vacuum angle $\theta$ that produces a dark neutron EDM $g$,
\al{
\sigma_{\bar{n}' p} \approx \epsilon^2 e^2 e^{\prime 2} g^2 v^2
\frac{m_p^4 m_{\bar{n}'}^2}{8\pi (m_p+m_{\bar{n}'})^4 m_{\gamma'}^4}.
}
Since we take pions as heavy as baryons, the quark masses are also approximately as heavy as baryons. Then, there is no chiral log enhancement \cite{Crewther:1979pi} and we use the tree-level result
\al{
g=-1.91 \frac{y m_{u'}m_{d'}}{\prn{m_{u'}+m_{d'}}m_{\bar{n}'}} \sin \theta \approx -1.91 \sin \theta.
}
Here, $y m_q$ is the quark mass contribution to the baryon mass and $y \approx 1.17$ in QCD. Note that the EDM dominates the magnetic dipole moment's contribution to scattering since the former is only suppressed by $v^2$ while the latter by $v^4$.

In addition to these direct detection signals, this ADM model may be discovered through its visibly decaying dark photons (necessary for entropy transfer after $\Lambda_{\text{QCD}'}$). Fig.~\ref{fig:darkphoton} shows the viable ranges of kinetic mixing and dark photon mass for the two possible DM scenarios described by Eqs.~\eqref{eq:mnp} and \eqref{eq:mpp}. The left panel corresponds to the case of $\bar{n}'$ DM where we have chosen $e'=10^{-2}e$ and the right panel to $\bar{p}'-\pi'^{+}$ DM with $e'=e_{\text{min}}$ (discussed below). For the dark photon parameter space with other interesting values of $e'$, see Appendix~\ref{sec:vissig_diffep}. In both scenarios, we take the limit that $\pi'$ is as heavy as the lightest dark baryon to allow heavier dark photons. The gray regions show currently excluded, visibly decaying dark photon parameter space, while the colored dashed lines show the projected sensitivities of future searches. 

The color shaded regions are obtained as follows. We require the dark photons as well as the neutral pions to decay before SM neutrinos decouple at $T_\nu^{\text{dec}}\sim 3 \text{ MeV}$~\cite{Mangano:2006ar} to avoid affecting $\Neff$ substantially. The first requirement places a lower bound on the kinetic mixing, shaded blue in Fig.~\ref{fig:darkphoton}. Below the SM pion threshold, the dark photon decay rate to a pair of leptons is simply
\al{
\Gamma_{\gamp \to \bar{l} l} = \frac{\alpha \epsilon^2 \prn{m_\gamp^2 + 2m_l^2}}{3 m_\gamp} \sqrt{1-\frac{4 m_l^2}{m_\gamp^2}}.
}
Above the pion threshold, we extract the decay rate using the measured ratios of cross sections with hadronic final states to those with muons~\cite{ParticleDataGroup:2020ssz}.

We further require that the dark photon is lighter than $\pi'^0$ to allow its decay. We assume that $m_{\bar{p}'}/f_{\pi'}$ is the same as in the SM when calculating the decay rate of $\pi'^{0}$. For dark photons with $m_\gamp \lesssim m_{\pi'}/2$, the minimum requisite dark electric charge to allow $\pi'^0 \to \gamp \gamp$ to proceed quickly enough is
\al{
e'_{\text{min}} = 4.5 \times 10^{-6}
}
for the case of $\bar{p}'-\pi'^{+}$ DM. In the region $m_{\pi'}/2 < m_\gamp < m_{\pi'}$, the neutral pion must instead decay through the vector portal to one dark and one SM photon. Therefore, its decay rate, given by
\al{
\Gamma_{\pi'^0 \to \gamp \gamma} = 2 \epsilon^2 \prn{1-\frac{m_\gamp^2}{m_{\pi'^0}^2}}^3 \frac{\alpha'^2}{64\pi^3} \prn{\frac{m_{\bar{p}'}}{f_{\pi'}}}^2 m_{\pi'^0},
}
is too slow for smaller $\epsilon$, as shaded orange in Fig.~\ref{fig:darkphoton}.

$e'_{\text{min}}$ is particularly interesting for the $\bar{p}'-\pi'^+$ DM case since it predicts an already-constrained direct detection cross section in some of the otherwise-viable dark photon parameter space. The current XENON1T direct detection bounds are the purple shaded regions in Fig.~\ref{fig:darkphoton}. As direct detection bounds improve in the near future, this minimum requisite dark electric charge will therefore complement proposed visibly-decaying dark photon searches and allow the $\bar{p}'-\pi'^{+}$ DM scenario to be constrained from two directions. 

Visibly decaying dark photons can be searched for in beam dump experiments or the Belle-II experiment, as seen in Fig.~\ref{fig:darkphoton}. When the dark photon is heavier than $\mathcal{O}(\text{GeV})$, however, an $e^+ e^-$ Higgs factory would be the best place to search for its decays to visible particles. It could also produce dark hadrons through an off-shell dark photon if $\epsilon$ is relatively large, in which case {\it dark spectroscopy}\/ could identify resonance states \cite{Hochberg:2017khi} in photon$+$missing signature, or with some dark states decaying into visible particles. In this case, it is in principle possible to confirm the $SU(3)$ gauge group with two flavors in the dark sector. The ILC can also accommodate beam dump experiment(s) reaching heavier dark photon masses than SHiP \cite{Kanemura:2015cxa}.

It is also quite likely that the dark Higgs and the standard model Higgs mix at some level through the quartic coupling $|H'|^2 |H|^2$ since no symmetry prohibits it. Then the 125~GeV Higgs can decay into dark neutrinos, dark photons, dark gluons, etc. Many of the states further decay back to the SM. Such exotic Higgs decays can be searched for particularly well at the $e^+ e^-$ Higgs factory, in some cases four orders of magnitude better than at the LHC \cite{Liu:2016zki}. 

\section{Discussion}
\label{sec:discuss}

In this paper, we have presented a simple dark sector which may be ``added'' onto existing models of electroweak baryogenesis to simultaneously explain DM. The dark sector contains a copy of the SM gauge group and at least one generation of matter with a right-handed neutrino and a Higgs doublet. The number of generations, Higgs doublets, and right-handed neutrinos may be varied, slightly affecting the final dark-sector asymmetries and therefore, the mass of DM. Throughout the main text, we have stuck to the minimal case of one of each. We have assumed that the mechanism of baryogenesis results in equal SM baryon and lepton asymmetries at the time of the EWPT. After, the neutrino portal allows some lepton asymmetry to transfer to the dark sector where it is converted into dark baryons which form (some of) DM. 

We have considered two simple benchmark scenarios: one in which $m_{u'} < m_{d'}$ such that this dark baryon asymmetry persists entirely in $\bar{p}'$, and one in which the opposite relation holds and the asymmetry is entirely in $\bar{n}'$. The former scenario has DM comprised of half $\bar{p}'$ and half $\pi'^{+}$, while in the latter, DM is entirely $\bar{n}'$. Both scenarios contain visibly decaying dark photons whose viable parameter space will be probed by current and future searches. Additionally, both may be probed by direct detection. In the former scenario in fact, the minimum requisite dark electric charge allows present direct detection bounds to already constrain a significant part of the otherwise-viable dark photon parameter space. A future $e^+ e^-$ Higgs factory can look for the possibly heavier dark photons in this model as well as exotic Higgs decays.

The parameter space of this model may be more fully explored by considering the case of finite, right-handed neutrino masses. For heavier masses, the asymmetry transfer from the SM to the dark sector becomes less efficient, decreasing the relative dark baryon asymmetry and increasing the DM mass. In the case of a strongly first order dark EWPT, there is also the possibility of gravitational waves from a low-scale (\emph{i.e.}, below electroweak) phase transition. The strongly interacting hadronic DM may also have sufficient self interactions to ameliorate known small-scale structure problems. All of these are interesting lines of future inquiry.

One way to view this model is as a simple ``stand in'' for any model of baryogenesis which otherwise lacks a natural DM candidate. Adding this dark sector to such a model will provide ADM without affecting the original mechanism of baryogenesis.

\acknowledgments

The work of EH was supported by the NSF GRFP.
RM is supported by the U.S. DoE under grant DE-SC0007859. RM thanks Michael Williams for providing the latest LHCb dark photon constraints.
HM was supported by the U.S. DoE under the Contract No. DE-AC02-
05CH11231, by the NSF grant PHY-1915314, by the JSPS grant JP20K03942,
MEXT Grant-in-Aid for Transformative Research Areas
(A) JP20H05850, JP20A203, by World Premier International
Research Center Initiative, MEXT, Japan, and
Hamamatsu Photonics, K.K. 
BS is supported by the UC Berkeley Chancellor's Fellowship throughout the year and over the summer of 2021 by the Professor Victor F. Lenzen Memorial Fund.

\appendix
\section{Visible Signals for Different \texorpdfstring{$e'$}{e'}}
\label{sec:vissig_diffep}

\begin{figure}[t!]
\includegraphics[width =\columnwidth]{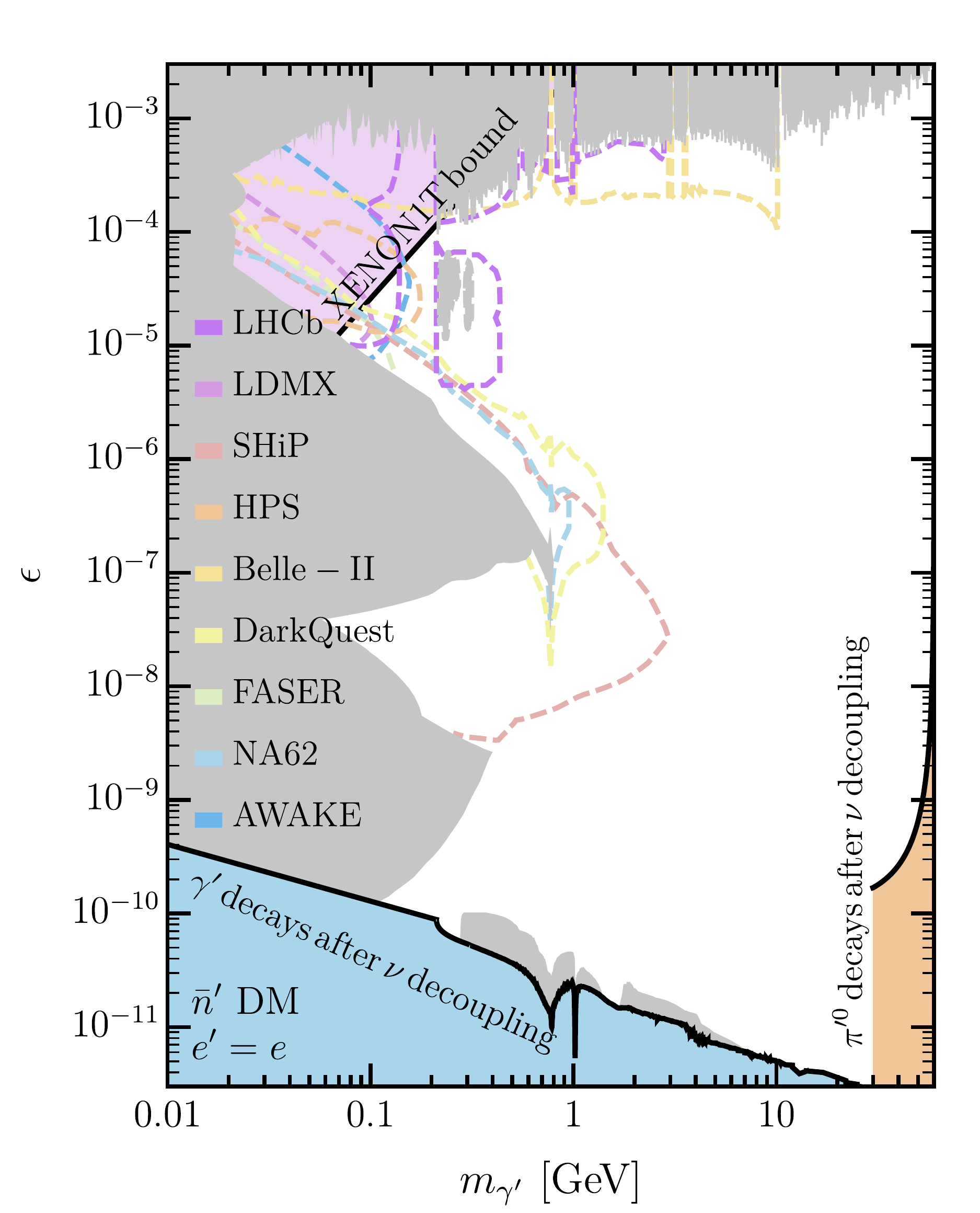}
\caption{\label{fig:np_DM_ep1} The same as the left panel of Fig.~\ref{fig:darkphoton}, with $e'= e$.} 
\end{figure}

\begin{figure*}[t!]
\includegraphics[width =\columnwidth]{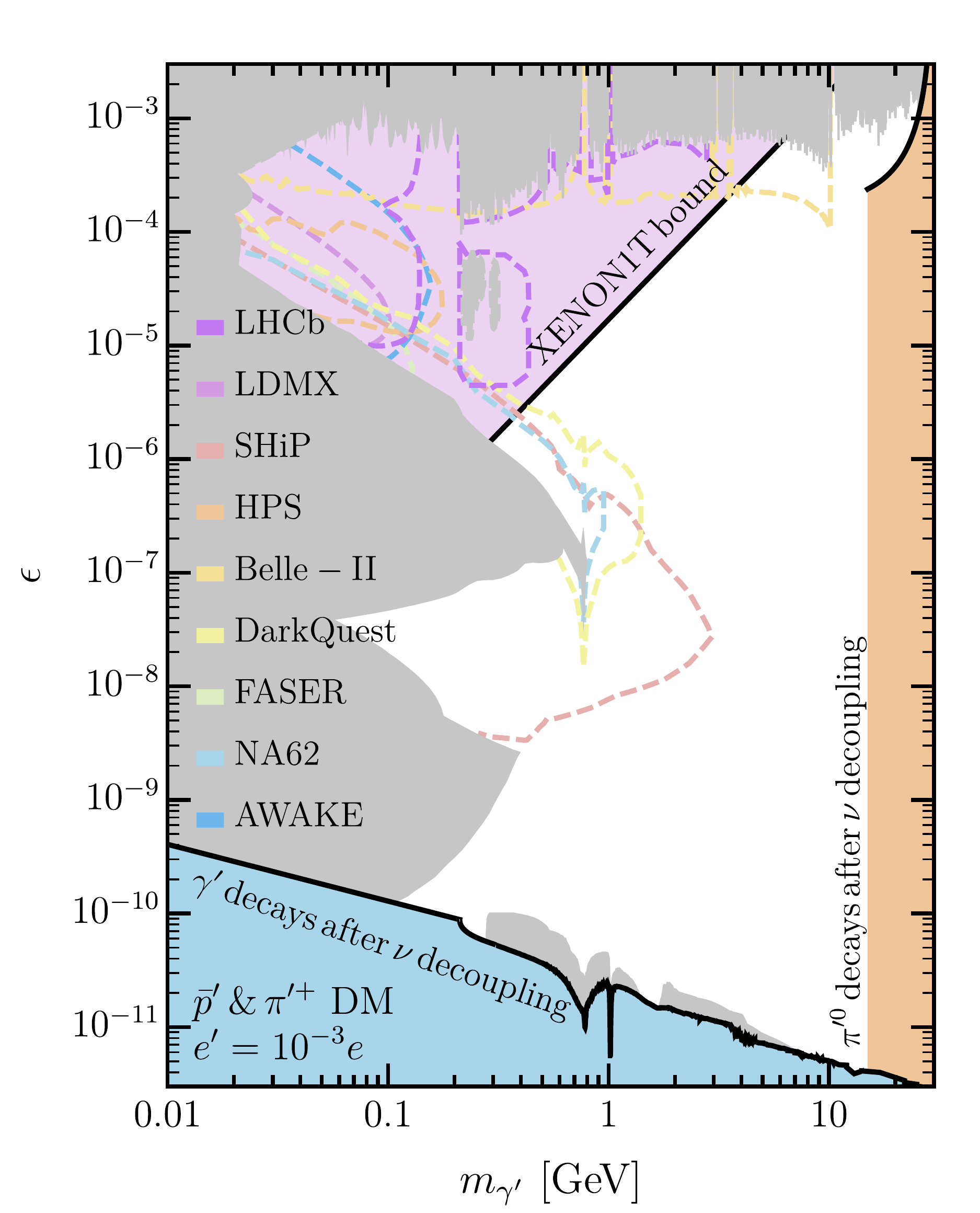}
\includegraphics[width =\columnwidth]{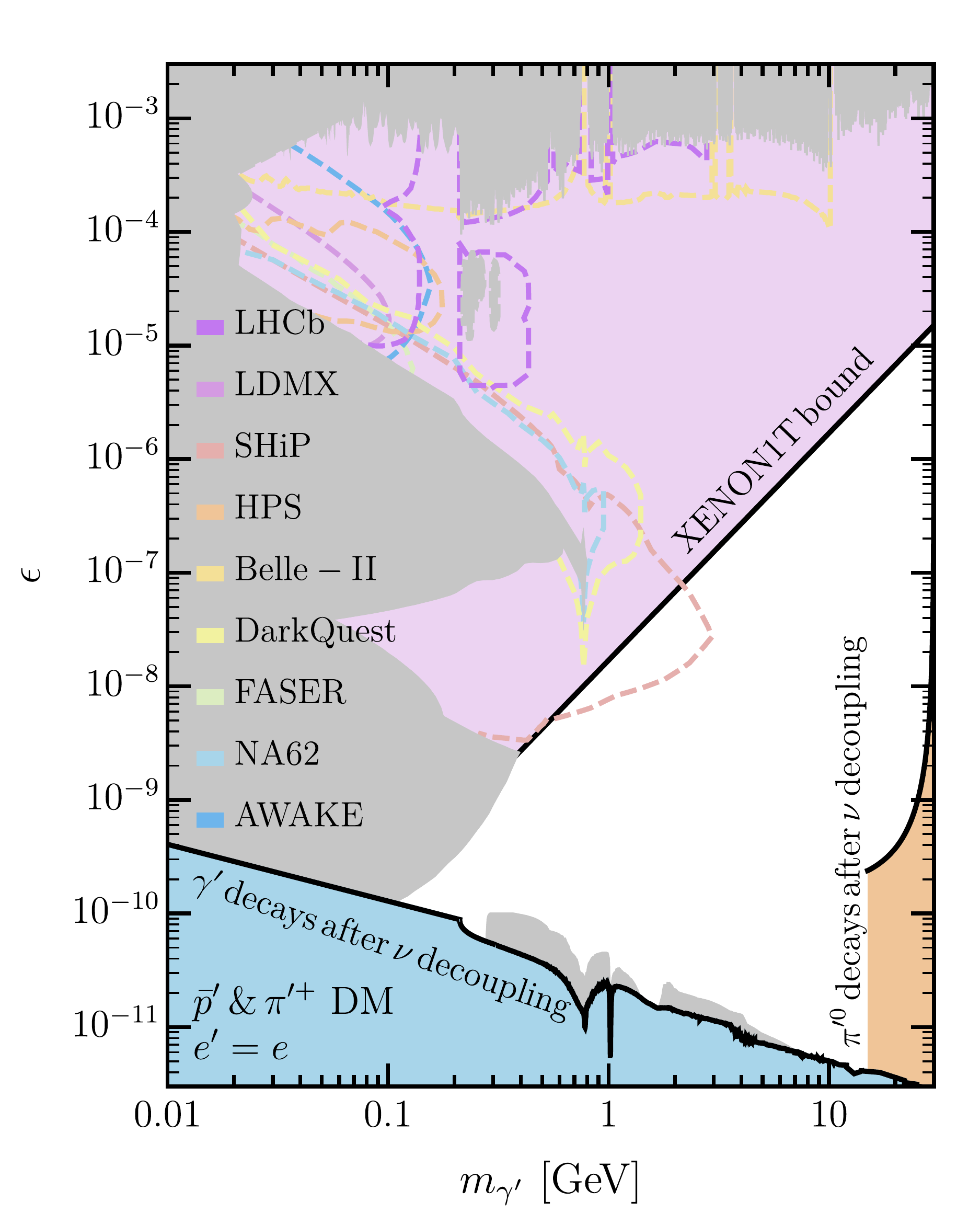}
\caption{\label{fig:pp_and_pip_DM} The same as the right panel of Fig.~\ref{fig:darkphoton}, with {\bf Left: } $e'= 10^{-3}e$ and {\bf Right: } $e'= e$.} 
\end{figure*}

Fig.~\ref{fig:np_DM_ep1} shows the viable dark photon parameter space for the $\bar{n}'$ DM case when $e'=e$, while Fig.~\ref{fig:pp_and_pip_DM} shows the $\bar{p}'-\pi'^+$ DM case for both $e'= 10^{-3}e$ and $e'= e$. All shaded regions, constraints, and projections correspond to those detailed in Fig.~\ref{fig:darkphoton}.

\section{General Lepton \& Baryon Asymmetries}
\label{sec:lepbarasym}
Although we enumerated the most minimal model for ADM coming from a SM baryogenesis scenario above, we present here the results for a more general dark sector. In particular, we expand the dark sector to include any number of dark generations, dark Higgs doublets, and dark right-handed neutrinos. We also permit an arbitrary initial lepton asymmetry produced during SM baryogenesis. In this more general scenario, Eq.~\eqref{eq:BLQ} for the baryons, leptons, and charges becomes
\begin{align}
    B &= 2 N (\mu_u + \mu_d) \nonumber \\
    B' &= 4 N' \mu'_{uL} + 2 N' \mu'_W \nonumber \\
    L &= 3N \mu_{N_R} + 2 N (\mu_d - \mu_u) \nonumber \\
   L' &= 3\mu' + 2 N' \mu'_W - N' \mu'_0 + N_{N_R} \mu_{N_R} \\
    Q &= (6N + 2m + 4) \mu_u - (4N + 2m + 4)\mu_d - 2 N \mu_{N_R} \nonumber \\
    Q' &= 2 N' \mu'_{uL} \! - \! 2 \mu' \! - \! (4 N'\! +\! 2 m'\! + 4) \mu'_W \! + \! (4 N' \! + \! 2 m') \mu'_0 \nonumber
\end{align}
where $N$ and $N'$ are numbers of generations in the SM and dark sector, $m$ and $m'$ are numbers of Higgs doublets in the SM and dark sector, and $N_{N_R}$ is the number of right-handed neutrinos. $\mu'$ is also now the sum over dark, left-handed neutrino chemical potentials.

Without loss of generality, we may assume that after the SM baryogenesis, $L$ begins at  $L_i = f B$ where $f$ can be any real, non-zero number. 
In lieu of Eq.~\eqref{eq:sphal}, while sphalerons and the neutrino portal in the dark sector are active, they impose
\al{
&3 N' \mu_{uL'}+2 N' \mu_{W'}+\mu'=0 & &
\mu_{0'}=\frac{\mu'}{N'}-\mu_{N_R}.
}

In this more general calculation, we again consider the two possibilities for the dark EWPT. If it is first order, we find that the baryon and lepton numbers take the form
\begin{align}
    B' &= -\frac{f_1(N, N', m, m', f)}{h_1(N_{N_R}, N, N', m, m')} B \nonumber \\
    L' &= \frac{g_1(N_{N_R}, N, N', m, m', f)}{h_1(N_{N_R}, N, N', m, m')} B 
\end{align}
where 
\begin{align}
    \lefteqn{
    f_1(N, N',m, m', f) } \nonumber\\ 
    &\phantom{=} 4N' \big[m' + 2 N'\big]\big[f (4+2m+5N) - N\big],
\end{align}
\begin{align}
    g_1(N_{N_R}, N, ...) =& 
    \big[3m'(N_{N_R}+3N')+2N'(N_{N_R}+7N')\big] \nonumber\\ 
    &\phantom{=}\times \big[f (4+2m+5N) - N\big],
\end{align}
\begin{align}
    \lefteqn{
    h_1(N_{N_R}, N, ...) } \nonumber \\
    &= 3m'\big[N(12+6m+11N)+N_{N_R}(4+2m+5N)\big] \nonumber \\
    &\phantom{=}+2N'\big[ 11N^2+2(2+m)(N_{N_R}+11N') \nonumber \\
    &\phantom{=+2N'\big[} +N(12+6m+5N_{N_R}+55N') \big] \nonumber \\
    &\phantom{=}+ 13m' N' (4+2m+5N).
\end{align}

If it is instead a crossover transition, we find them to be
\begin{align}
    B' &= -\frac{f_2(N, N', m, m', f)}{h_2(N_{N_R}, N, N', m, m')} B, \nonumber \\
    L' &= \frac{g_2(N_{N_R}, N, N', m, m', f)}{h_2(N_{N_R}, N, N', m, m')} B,
\end{align}
where 
\begin{align}
    \lefteqn{
    f_2(N, N', m, m', f) } \nonumber\\ 
    &= 4N' \big[m' + 2 N' + 2\big]\big[f (2 m + 5 N + 4) - N\big],
\end{align}
\begin{align}
    \lefteqn{
    g_2(N_{N_R}, N, ...) } \nonumber \\ 
    &= \big[ N_{N_R} (6+3m'+8N')+N'(18+9m'+16N') \big] \nonumber\\ 
    &\phantom{=}\times \big[f (2 m + 5 N + 4) - N\big],
\end{align}
\begin{align}
\lefteqn{
    h_2(N_{N_R}, N, ...) } \nonumber \\
    &= N \big[ 3(2+m')(12+6m+5N_{N_R})+120N'^2 \nonumber \\
    &\phantom{=N \big[}+(226+48m+65m'+40N_{N_R})N'\big] \nonumber \\
    &\phantom{=}+2(2+m)\big[ N_{N_R}(6+3m'+8N') \nonumber \\
    &\phantom{=+2(2+m)\big[}+N'(26+13m'+24N') \big] \nonumber \\
    &\phantom{=} +11N^2(6+3m'+8N')
\end{align}

If $m_{d'}<m_{u'}$ so that DM is dark anti-neutrons, its mass is
\begin{align}
    \label{eq:genmn'}
    m_{\bar{n}'} &= \frac{m_n + 7m_p}{8} \frac{|B|}{|B'|}\frac{\Omega_{DM}}{\Omega_{SM}},
\end{align}
where $m_{n}$ and $m_p$ are the masses of the SM neutron and proton and $\Omega_{DM}$ and $\Omega_{SM}$ are the energy densities of DM and baryons~\cite{Zyla:2020zbs}. If instead $m_{u'}<m_{d'}$ so that DM is equal numbers of dark anti-protons and pions, then their masses satisfy:
\begin{align}
    \label{eq:genmp'}
    m_{\bar{p}'} + m_{\pi} &= \frac{m_n + 7m_p}{8} \frac{|B|}{|B'|}\frac{\Omega_{DM}}{\Omega_{SM}}.
\end{align}

We list below the asymmetries and masses for several interesting, but less minimal, dark sectors. For a first order dark EWPT where $N' = N_{N_R} = m' = 1$, we find the asymmetries to be: 
\begin{equation}
    B' = -\frac{72}{535}B, \hspace{5 mm} L' = \frac{168}{535}B,
\end{equation}
which lead to a predicted dark anti-neutron mass of 
\begin{equation}
m_{\bar{n}'} = 37.4 \text{ GeV}
\end{equation}
or a predicted dark anti-proton and pion mass of
\begin{equation}
m_{\bar{p}'} = m_{\pi} = 18.7 \text{ GeV}.
\end{equation}
This dark sector is the same as the one considered in the main text, except that its EWPT is first order instead of second.

Another interesting dark sector, motivated by the SM, is when $N' = N_{N_R} = 3$ and $m' = 1$. For a crossover dark EWPT, we find the predicted asymmetries and masses to be:
\begin{equation}
    B' = -\frac{72}{523}B, \hspace{5 mm} L' = \frac{216}{523}B,
\end{equation}
and 
\begin{equation}
    m_{\bar{n}'} = 36.6 \text{ GeV} \hspace{3 mm}\text{or}\hspace{3 mm} m_{\bar{p}'} = m_{\pi} = 18.3 \text{ GeV}.
\end{equation}
On the other hand, for the dark first order phase transition, we find them to be:
\begin{equation}
    B' = -\frac{168}{769}B, \hspace{5 mm} L' = \frac{360}{769}B,
\end{equation}
and 
\begin{equation}
    m_{\bar{n}'} = 23.0 \text{ GeV} \hspace{3 mm}\text{or}\hspace{3 mm} m_{\bar{p}'} = m_{\pi} = 11.5 \text{ GeV}.
\end{equation}

\bibliography{Refs.bib}

\begin{thebibliography}{58}%
\makeatletter
\providecommand \@ifxundefined [1]{%
 \@ifx{#1\undefined}
}%
\providecommand \@ifnum [1]{%
 \ifnum #1\expandafter \@firstoftwo
 \else \expandafter \@secondoftwo
 \fi
}%
\providecommand \@ifx [1]{%
 \ifx #1\expandafter \@firstoftwo
 \else \expandafter \@secondoftwo
 \fi
}%
\providecommand \natexlab [1]{#1}%
\providecommand \enquote  [1]{``#1''}%
\providecommand \bibnamefont  [1]{#1}%
\providecommand \bibfnamefont [1]{#1}%
\providecommand \citenamefont [1]{#1}%
\providecommand \href@noop [0]{\@secondoftwo}%
\providecommand \href [0]{\begingroup \@sanitize@url \@href}%
\providecommand \@href[1]{\@@startlink{#1}\@@href}%
\providecommand \@@href[1]{\endgroup#1\@@endlink}%
\providecommand \@sanitize@url [0]{\catcode `\\12\catcode `\$12\catcode
  `\&12\catcode `\#12\catcode `\^12\catcode `\_12\catcode `\%12\relax}%
\providecommand \@@startlink[1]{}%
\providecommand \@@endlink[0]{}%
\providecommand \url  [0]{\begingroup\@sanitize@url \@url }%
\providecommand \@url [1]{\endgroup\@href {#1}{\urlprefix }}%
\providecommand \urlprefix  [0]{URL }%
\providecommand \Eprint [0]{\href }%
\providecommand \doibase [0]{http://dx.doi.org/}%
\providecommand \selectlanguage [0]{\@gobble}%
\providecommand \bibinfo  [0]{\@secondoftwo}%
\providecommand \bibfield  [0]{\@secondoftwo}%
\providecommand \translation [1]{[#1]}%
\providecommand \BibitemOpen [0]{}%
\providecommand \bibitemStop [0]{}%
\providecommand \bibitemNoStop [0]{.\EOS\space}%
\providecommand \EOS [0]{\spacefactor3000\relax}%
\providecommand \BibitemShut  [1]{\csname bibitem#1\endcsname}%
\let\auto@bib@innerbib\@empty
\bibitem [{\citenamefont {Aprile}\ \emph {et~al.}(2018)\citenamefont {Aprile}
  \emph {et~al.}}]{Aprile:2018dbl}%
  \BibitemOpen
  \bibfield  {author} {\bibinfo {author} {\bibfnamefont {E.}~\bibnamefont
  {Aprile}} \emph {et~al.} (\bibinfo {collaboration} {XENON}),\ }\href
  {\doibase 10.1103/PhysRevLett.121.111302} {\bibfield  {journal} {\bibinfo
  {journal} {Phys. Rev. Lett.}\ }\textbf {\bibinfo {volume} {121}},\ \bibinfo
  {pages} {111302} (\bibinfo {year} {2018})},\ \Eprint
  {http://arxiv.org/abs/1805.12562} {arXiv:1805.12562 [astro-ph.CO]}
  \BibitemShut {NoStop}%
\bibitem [{\citenamefont {Leane}\ \emph {et~al.}(2018)\citenamefont {Leane},
  \citenamefont {Slatyer}, \citenamefont {Beacom},\ and\ \citenamefont
  {Ng}}]{Leane:2018kjk}%
  \BibitemOpen
  \bibfield  {author} {\bibinfo {author} {\bibfnamefont {R.~K.}\ \bibnamefont
  {Leane}}, \bibinfo {author} {\bibfnamefont {T.~R.}\ \bibnamefont {Slatyer}},
  \bibinfo {author} {\bibfnamefont {J.~F.}\ \bibnamefont {Beacom}}, \ and\
  \bibinfo {author} {\bibfnamefont {K.~C.}\ \bibnamefont {Ng}},\ }\href
  {\doibase 10.1103/PhysRevD.98.023016} {\bibfield  {journal} {\bibinfo
  {journal} {Phys. Rev. D}\ }\textbf {\bibinfo {volume} {98}},\ \bibinfo
  {pages} {023016} (\bibinfo {year} {2018})},\ \Eprint
  {http://arxiv.org/abs/1805.10305} {arXiv:1805.10305 [hep-ph]} \BibitemShut
  {NoStop}%
\bibitem [{\citenamefont {Jarlskog}(1985)}]{Jarlskog:1985ht}%
  \BibitemOpen
  \bibfield  {author} {\bibinfo {author} {\bibfnamefont {C.}~\bibnamefont
  {Jarlskog}},\ }\href {\doibase 10.1103/PhysRevLett.55.1039} {\bibfield
  {journal} {\bibinfo  {journal} {Phys. Rev. Lett.}\ }\textbf {\bibinfo
  {volume} {55}},\ \bibinfo {pages} {1039} (\bibinfo {year}
  {1985})}\BibitemShut {NoStop}%
\bibitem [{\citenamefont {Gavela}\ \emph {et~al.}(1994)\citenamefont {Gavela},
  \citenamefont {Hernandez}, \citenamefont {Orloff}, \citenamefont {Pene},\
  and\ \citenamefont {Quimbay}}]{Gavela:1994dt}%
  \BibitemOpen
  \bibfield  {author} {\bibinfo {author} {\bibfnamefont {M.~B.}\ \bibnamefont
  {Gavela}}, \bibinfo {author} {\bibfnamefont {P.}~\bibnamefont {Hernandez}},
  \bibinfo {author} {\bibfnamefont {J.}~\bibnamefont {Orloff}}, \bibinfo
  {author} {\bibfnamefont {O.}~\bibnamefont {Pene}}, \ and\ \bibinfo {author}
  {\bibfnamefont {C.}~\bibnamefont {Quimbay}},\ }\href {\doibase
  10.1016/0550-3213(94)00410-2} {\bibfield  {journal} {\bibinfo  {journal}
  {Nucl. Phys. B}\ }\textbf {\bibinfo {volume} {430}},\ \bibinfo {pages} {382}
  (\bibinfo {year} {1994})},\ \Eprint {http://arxiv.org/abs/hep-ph/9406289}
  {arXiv:hep-ph/9406289} \BibitemShut {NoStop}%
\bibitem [{\citenamefont {Huet}\ and\ \citenamefont
  {Sather}(1995)}]{Huet:1994jb}%
  \BibitemOpen
  \bibfield  {author} {\bibinfo {author} {\bibfnamefont {P.}~\bibnamefont
  {Huet}}\ and\ \bibinfo {author} {\bibfnamefont {E.}~\bibnamefont {Sather}},\
  }\href {\doibase 10.1103/PhysRevD.51.379} {\bibfield  {journal} {\bibinfo
  {journal} {Phys. Rev. D}\ }\textbf {\bibinfo {volume} {51}},\ \bibinfo
  {pages} {379} (\bibinfo {year} {1995})},\ \Eprint
  {http://arxiv.org/abs/hep-ph/9404302} {arXiv:hep-ph/9404302} \BibitemShut
  {NoStop}%
\bibitem [{\citenamefont {Kajantie}\ \emph {et~al.}(1996)\citenamefont
  {Kajantie}, \citenamefont {Laine}, \citenamefont {Rummukainen},\ and\
  \citenamefont {Shaposhnikov}}]{Kajantie:1995kf}%
  \BibitemOpen
  \bibfield  {author} {\bibinfo {author} {\bibfnamefont {K.}~\bibnamefont
  {Kajantie}}, \bibinfo {author} {\bibfnamefont {M.}~\bibnamefont {Laine}},
  \bibinfo {author} {\bibfnamefont {K.}~\bibnamefont {Rummukainen}}, \ and\
  \bibinfo {author} {\bibfnamefont {M.~E.}\ \bibnamefont {Shaposhnikov}},\
  }\href {\doibase 10.1016/0550-3213(96)00052-1} {\bibfield  {journal}
  {\bibinfo  {journal} {Nucl. Phys. B}\ }\textbf {\bibinfo {volume} {466}},\
  \bibinfo {pages} {189} (\bibinfo {year} {1996})},\ \Eprint
  {http://arxiv.org/abs/hep-lat/9510020} {arXiv:hep-lat/9510020} \BibitemShut
  {NoStop}%
\bibitem [{\citenamefont {Kajantie}\ \emph {et~al.}(1997)\citenamefont
  {Kajantie}, \citenamefont {Laine}, \citenamefont {Rummukainen},\ and\
  \citenamefont {Shaposhnikov}}]{Kajantie:1996qd}%
  \BibitemOpen
  \bibfield  {author} {\bibinfo {author} {\bibfnamefont {K.}~\bibnamefont
  {Kajantie}}, \bibinfo {author} {\bibfnamefont {M.}~\bibnamefont {Laine}},
  \bibinfo {author} {\bibfnamefont {K.}~\bibnamefont {Rummukainen}}, \ and\
  \bibinfo {author} {\bibfnamefont {M.~E.}\ \bibnamefont {Shaposhnikov}},\
  }\href {\doibase 10.1016/S0550-3213(97)00164-8} {\bibfield  {journal}
  {\bibinfo  {journal} {Nucl. Phys. B}\ }\textbf {\bibinfo {volume} {493}},\
  \bibinfo {pages} {413} (\bibinfo {year} {1997})},\ \Eprint
  {http://arxiv.org/abs/hep-lat/9612006} {arXiv:hep-lat/9612006} \BibitemShut
  {NoStop}%
\bibitem [{\citenamefont {Rummukainen}\ \emph {et~al.}(1998)\citenamefont
  {Rummukainen}, \citenamefont {Tsypin}, \citenamefont {Kajantie},
  \citenamefont {Laine},\ and\ \citenamefont
  {Shaposhnikov}}]{Rummukainen:1998as}%
  \BibitemOpen
  \bibfield  {author} {\bibinfo {author} {\bibfnamefont {K.}~\bibnamefont
  {Rummukainen}}, \bibinfo {author} {\bibfnamefont {M.}~\bibnamefont {Tsypin}},
  \bibinfo {author} {\bibfnamefont {K.}~\bibnamefont {Kajantie}}, \bibinfo
  {author} {\bibfnamefont {M.}~\bibnamefont {Laine}}, \ and\ \bibinfo {author}
  {\bibfnamefont {M.~E.}\ \bibnamefont {Shaposhnikov}},\ }\href {\doibase
  10.1016/S0550-3213(98)00494-5} {\bibfield  {journal} {\bibinfo  {journal}
  {Nucl. Phys. B}\ }\textbf {\bibinfo {volume} {532}},\ \bibinfo {pages} {283}
  (\bibinfo {year} {1998})},\ \Eprint {http://arxiv.org/abs/hep-lat/9805013}
  {arXiv:hep-lat/9805013} \BibitemShut {NoStop}%
\bibitem [{\citenamefont {Fukugita}\ and\ \citenamefont
  {Yanagida}(1986)}]{Fukugita:1986hr}%
  \BibitemOpen
  \bibfield  {author} {\bibinfo {author} {\bibfnamefont {M.}~\bibnamefont
  {Fukugita}}\ and\ \bibinfo {author} {\bibfnamefont {T.}~\bibnamefont
  {Yanagida}},\ }\href {\doibase 10.1016/0370-2693(86)91126-3} {\bibfield
  {journal} {\bibinfo  {journal} {Phys. Lett. B}\ }\textbf {\bibinfo {volume}
  {174}},\ \bibinfo {pages} {45} (\bibinfo {year} {1986})}\BibitemShut
  {NoStop}%
\bibitem [{\citenamefont {Cohen}\ \emph {et~al.}(1993)\citenamefont {Cohen},
  \citenamefont {Kaplan},\ and\ \citenamefont {Nelson}}]{Cohen:1993nk}%
  \BibitemOpen
  \bibfield  {author} {\bibinfo {author} {\bibfnamefont {A.~G.}\ \bibnamefont
  {Cohen}}, \bibinfo {author} {\bibfnamefont {D.~B.}\ \bibnamefont {Kaplan}}, \
  and\ \bibinfo {author} {\bibfnamefont {A.~E.}\ \bibnamefont {Nelson}},\
  }\href {\doibase 10.1146/annurev.ns.43.120193.000331} {\bibfield  {journal}
  {\bibinfo  {journal} {Ann. Rev. Nucl. Part. Sci.}\ }\textbf {\bibinfo
  {volume} {43}},\ \bibinfo {pages} {27} (\bibinfo {year} {1993})},\ \Eprint
  {http://arxiv.org/abs/hep-ph/9302210} {arXiv:hep-ph/9302210} \BibitemShut
  {NoStop}%
\bibitem [{\citenamefont {Morrissey}\ and\ \citenamefont
  {Ramsey-Musolf}(2012)}]{Morrissey:2012db}%
  \BibitemOpen
  \bibfield  {author} {\bibinfo {author} {\bibfnamefont {D.~E.}\ \bibnamefont
  {Morrissey}}\ and\ \bibinfo {author} {\bibfnamefont {M.~J.}\ \bibnamefont
  {Ramsey-Musolf}},\ }\href {\doibase 10.1088/1367-2630/14/12/125003}
  {\bibfield  {journal} {\bibinfo  {journal} {New J. Phys.}\ }\textbf {\bibinfo
  {volume} {14}},\ \bibinfo {pages} {125003} (\bibinfo {year} {2012})},\
  \Eprint {http://arxiv.org/abs/1206.2942} {arXiv:1206.2942 [hep-ph]}
  \BibitemShut {NoStop}%
\bibitem [{\citenamefont {Konstandin}(2013)}]{Konstandin:2013caa}%
  \BibitemOpen
  \bibfield  {author} {\bibinfo {author} {\bibfnamefont {T.}~\bibnamefont
  {Konstandin}},\ }\href {\doibase 10.3367/UFNe.0183.201308a.0785} {\bibfield
  {journal} {\bibinfo  {journal} {Phys. Usp.}\ }\textbf {\bibinfo {volume}
  {56}},\ \bibinfo {pages} {747} (\bibinfo {year} {2013})},\ \Eprint
  {http://arxiv.org/abs/1302.6713} {arXiv:1302.6713 [hep-ph]} \BibitemShut
  {NoStop}%
\bibitem [{\citenamefont {Elor}\ \emph {et~al.}(2019)\citenamefont {Elor},
  \citenamefont {Escudero},\ and\ \citenamefont {Nelson}}]{Elor:2018twp}%
  \BibitemOpen
  \bibfield  {author} {\bibinfo {author} {\bibfnamefont {G.}~\bibnamefont
  {Elor}}, \bibinfo {author} {\bibfnamefont {M.}~\bibnamefont {Escudero}}, \
  and\ \bibinfo {author} {\bibfnamefont {A.}~\bibnamefont {Nelson}},\ }\href
  {\doibase 10.1103/PhysRevD.99.035031} {\bibfield  {journal} {\bibinfo
  {journal} {Phys. Rev. D}\ }\textbf {\bibinfo {volume} {99}},\ \bibinfo
  {pages} {035031} (\bibinfo {year} {2019})},\ \Eprint
  {http://arxiv.org/abs/1810.00880} {arXiv:1810.00880 [hep-ph]} \BibitemShut
  {NoStop}%
\bibitem [{\citenamefont {Alonso-\'Alvarez}\ \emph {et~al.}(2020)\citenamefont
  {Alonso-\'Alvarez}, \citenamefont {Elor}, \citenamefont {Nelson},\ and\
  \citenamefont {Xiao}}]{Alonso-Alvarez:2019fym}%
  \BibitemOpen
  \bibfield  {author} {\bibinfo {author} {\bibfnamefont {G.}~\bibnamefont
  {Alonso-\'Alvarez}}, \bibinfo {author} {\bibfnamefont {G.}~\bibnamefont
  {Elor}}, \bibinfo {author} {\bibfnamefont {A.~E.}\ \bibnamefont {Nelson}}, \
  and\ \bibinfo {author} {\bibfnamefont {H.}~\bibnamefont {Xiao}},\ }\href
  {\doibase 10.1007/JHEP03(2020)046} {\bibfield  {journal} {\bibinfo  {journal}
  {JHEP}\ }\textbf {\bibinfo {volume} {03}},\ \bibinfo {pages} {046} (\bibinfo
  {year} {2020})},\ \Eprint {http://arxiv.org/abs/1907.10612} {arXiv:1907.10612
  [hep-ph]} \BibitemShut {NoStop}%
\bibitem [{\citenamefont {Elor}\ and\ \citenamefont
  {McGehee}(2021)}]{Elor:2020tkc}%
  \BibitemOpen
  \bibfield  {author} {\bibinfo {author} {\bibfnamefont {G.}~\bibnamefont
  {Elor}}\ and\ \bibinfo {author} {\bibfnamefont {R.}~\bibnamefont {McGehee}},\
  }\href {\doibase 10.1103/PhysRevD.103.035005} {\bibfield  {journal} {\bibinfo
   {journal} {Phys. Rev. D}\ }\textbf {\bibinfo {volume} {103}},\ \bibinfo
  {pages} {035005} (\bibinfo {year} {2021})},\ \Eprint
  {http://arxiv.org/abs/2011.06115} {arXiv:2011.06115 [hep-ph]} \BibitemShut
  {NoStop}%
\bibitem [{\citenamefont {Minkowski}(1977)}]{Minkowski:1977sc}%
  \BibitemOpen
  \bibfield  {author} {\bibinfo {author} {\bibfnamefont {P.}~\bibnamefont
  {Minkowski}},\ }\href {\doibase 10.1016/0370-2693(77)90435-X} {\bibfield
  {journal} {\bibinfo  {journal} {Phys. Lett. B}\ }\textbf {\bibinfo {volume}
  {67}},\ \bibinfo {pages} {421} (\bibinfo {year} {1977})}\BibitemShut
  {NoStop}%
\bibitem [{\citenamefont {Gell-Mann}\ \emph {et~al.}(1979)\citenamefont
  {Gell-Mann}, \citenamefont {Ramond},\ and\ \citenamefont
  {Slansky}}]{GellMann:1980vs}%
  \BibitemOpen
  \bibfield  {author} {\bibinfo {author} {\bibfnamefont {M.}~\bibnamefont
  {Gell-Mann}}, \bibinfo {author} {\bibfnamefont {P.}~\bibnamefont {Ramond}}, \
  and\ \bibinfo {author} {\bibfnamefont {R.}~\bibnamefont {Slansky}},\
  }\href@noop {} {\bibfield  {journal} {\bibinfo  {journal} {Conf. Proc. C}\
  }\textbf {\bibinfo {volume} {790927}},\ \bibinfo {pages} {315} (\bibinfo
  {year} {1979})},\ \Eprint {http://arxiv.org/abs/1306.4669} {arXiv:1306.4669
  [hep-th]} \BibitemShut {NoStop}%
\bibitem [{\citenamefont {Yanagida}(1980)}]{Yanagida:1980xy}%
  \BibitemOpen
  \bibfield  {author} {\bibinfo {author} {\bibfnamefont {T.}~\bibnamefont
  {Yanagida}},\ }\href {\doibase 10.1143/PTP.64.1103} {\bibfield  {journal}
  {\bibinfo  {journal} {Prog. Theor. Phys.}\ }\textbf {\bibinfo {volume}
  {64}},\ \bibinfo {pages} {1103} (\bibinfo {year} {1980})}\BibitemShut
  {NoStop}%
\bibitem [{\citenamefont {Buchmuller}\ \emph {et~al.}(2005)\citenamefont
  {Buchmuller}, \citenamefont {Di~Bari},\ and\ \citenamefont
  {Plumacher}}]{Buchmuller:2004nz}%
  \BibitemOpen
  \bibfield  {author} {\bibinfo {author} {\bibfnamefont {W.}~\bibnamefont
  {Buchmuller}}, \bibinfo {author} {\bibfnamefont {P.}~\bibnamefont {Di~Bari}},
  \ and\ \bibinfo {author} {\bibfnamefont {M.}~\bibnamefont {Plumacher}},\
  }\href {\doibase 10.1016/j.aop.2004.02.003} {\bibfield  {journal} {\bibinfo
  {journal} {Annals Phys.}\ }\textbf {\bibinfo {volume} {315}},\ \bibinfo
  {pages} {305} (\bibinfo {year} {2005})},\ \Eprint
  {http://arxiv.org/abs/hep-ph/0401240} {arXiv:hep-ph/0401240} \BibitemShut
  {NoStop}%
\bibitem [{\citenamefont {Dror}\ \emph {et~al.}(2020)\citenamefont {Dror},
  \citenamefont {Hiramatsu}, \citenamefont {Kohri}, \citenamefont {Murayama},\
  and\ \citenamefont {White}}]{Dror:2019syi}%
  \BibitemOpen
  \bibfield  {author} {\bibinfo {author} {\bibfnamefont {J.~A.}\ \bibnamefont
  {Dror}}, \bibinfo {author} {\bibfnamefont {T.}~\bibnamefont {Hiramatsu}},
  \bibinfo {author} {\bibfnamefont {K.}~\bibnamefont {Kohri}}, \bibinfo
  {author} {\bibfnamefont {H.}~\bibnamefont {Murayama}}, \ and\ \bibinfo
  {author} {\bibfnamefont {G.}~\bibnamefont {White}},\ }\href {\doibase
  10.1103/PhysRevLett.124.041804} {\bibfield  {journal} {\bibinfo  {journal}
  {Phys. Rev. Lett.}\ }\textbf {\bibinfo {volume} {124}},\ \bibinfo {pages}
  {041804} (\bibinfo {year} {2020})},\ \Eprint
  {http://arxiv.org/abs/1908.03227} {arXiv:1908.03227 [hep-ph]} \BibitemShut
  {NoStop}%
\bibitem [{\citenamefont {Andreev}\ \emph {et~al.}(2018)\citenamefont {Andreev}
  \emph {et~al.}}]{Andreev:2018ayy}%
  \BibitemOpen
  \bibfield  {author} {\bibinfo {author} {\bibfnamefont {V.}~\bibnamefont
  {Andreev}} \emph {et~al.} (\bibinfo {collaboration} {ACME}),\ }\href
  {\doibase 10.1038/s41586-018-0599-8} {\bibfield  {journal} {\bibinfo
  {journal} {Nature}\ }\textbf {\bibinfo {volume} {562}},\ \bibinfo {pages}
  {355} (\bibinfo {year} {2018})}\BibitemShut {NoStop}%
\bibitem [{\citenamefont {Demidov}\ \emph {et~al.}(2016)\citenamefont
  {Demidov}, \citenamefont {Gorbunov},\ and\ \citenamefont
  {Kirpichnikov}}]{Demidov:2016wcv}%
  \BibitemOpen
  \bibfield  {author} {\bibinfo {author} {\bibfnamefont {S.~V.}\ \bibnamefont
  {Demidov}}, \bibinfo {author} {\bibfnamefont {D.~S.}\ \bibnamefont
  {Gorbunov}}, \ and\ \bibinfo {author} {\bibfnamefont {D.~V.}\ \bibnamefont
  {Kirpichnikov}},\ }\href {\doibase 10.1007/JHEP11(2016)148} {\bibfield
  {journal} {\bibinfo  {journal} {JHEP}\ }\textbf {\bibinfo {volume} {11}},\
  \bibinfo {pages} {148} (\bibinfo {year} {2016})},\ \bibinfo {note} {[Erratum:
  JHEP 08, 080 (2017)]},\ \Eprint {http://arxiv.org/abs/1608.01985}
  {arXiv:1608.01985 [hep-ph]} \BibitemShut {NoStop}%
\bibitem [{\citenamefont {Dorsch}\ \emph {et~al.}(2017)\citenamefont {Dorsch},
  \citenamefont {Huber}, \citenamefont {Konstandin},\ and\ \citenamefont
  {No}}]{Dorsch:2016nrg}%
  \BibitemOpen
  \bibfield  {author} {\bibinfo {author} {\bibfnamefont {G.~C.}\ \bibnamefont
  {Dorsch}}, \bibinfo {author} {\bibfnamefont {S.~J.}\ \bibnamefont {Huber}},
  \bibinfo {author} {\bibfnamefont {T.}~\bibnamefont {Konstandin}}, \ and\
  \bibinfo {author} {\bibfnamefont {J.~M.}\ \bibnamefont {No}},\ }\href
  {\doibase 10.1088/1475-7516/2017/05/052} {\bibfield  {journal} {\bibinfo
  {journal} {JCAP}\ }\textbf {\bibinfo {volume} {05}},\ \bibinfo {pages} {052}
  (\bibinfo {year} {2017})},\ \Eprint {http://arxiv.org/abs/1611.05874}
  {arXiv:1611.05874 [hep-ph]} \BibitemShut {NoStop}%
\bibitem [{\citenamefont {Vaskonen}(2017)}]{Vaskonen:2016yiu}%
  \BibitemOpen
  \bibfield  {author} {\bibinfo {author} {\bibfnamefont {V.}~\bibnamefont
  {Vaskonen}},\ }\href {\doibase 10.1103/PhysRevD.95.123515} {\bibfield
  {journal} {\bibinfo  {journal} {Phys. Rev. D}\ }\textbf {\bibinfo {volume}
  {95}},\ \bibinfo {pages} {123515} (\bibinfo {year} {2017})},\ \Eprint
  {http://arxiv.org/abs/1611.02073} {arXiv:1611.02073 [hep-ph]} \BibitemShut
  {NoStop}%
\bibitem [{\citenamefont {Bruggisser}\ \emph {et~al.}(2018)\citenamefont
  {Bruggisser}, \citenamefont {Von~Harling}, \citenamefont {Matsedonskyi},\
  and\ \citenamefont {Servant}}]{Bruggisser:2018mrt}%
  \BibitemOpen
  \bibfield  {author} {\bibinfo {author} {\bibfnamefont {S.}~\bibnamefont
  {Bruggisser}}, \bibinfo {author} {\bibfnamefont {B.}~\bibnamefont
  {Von~Harling}}, \bibinfo {author} {\bibfnamefont {O.}~\bibnamefont
  {Matsedonskyi}}, \ and\ \bibinfo {author} {\bibfnamefont {G.}~\bibnamefont
  {Servant}},\ }\href {\doibase 10.1007/JHEP12(2018)099} {\bibfield  {journal}
  {\bibinfo  {journal} {JHEP}\ }\textbf {\bibinfo {volume} {12}},\ \bibinfo
  {pages} {099} (\bibinfo {year} {2018})},\ \Eprint
  {http://arxiv.org/abs/1804.07314} {arXiv:1804.07314 [hep-ph]} \BibitemShut
  {NoStop}%
\bibitem [{\citenamefont {Hall}\ \emph {et~al.}(2020)\citenamefont {Hall},
  \citenamefont {Konstandin}, \citenamefont {McGehee}, \citenamefont
  {Murayama},\ and\ \citenamefont {Servant}}]{Hall:2019ank}%
  \BibitemOpen
  \bibfield  {author} {\bibinfo {author} {\bibfnamefont {E.}~\bibnamefont
  {Hall}}, \bibinfo {author} {\bibfnamefont {T.}~\bibnamefont {Konstandin}},
  \bibinfo {author} {\bibfnamefont {R.}~\bibnamefont {McGehee}}, \bibinfo
  {author} {\bibfnamefont {H.}~\bibnamefont {Murayama}}, \ and\ \bibinfo
  {author} {\bibfnamefont {G.}~\bibnamefont {Servant}},\ }\href {\doibase
  10.1007/JHEP04(2020)042} {\bibfield  {journal} {\bibinfo  {journal} {JHEP}\
  }\textbf {\bibinfo {volume} {04}},\ \bibinfo {pages} {042} (\bibinfo {year}
  {2020})},\ \Eprint {http://arxiv.org/abs/1910.08068} {arXiv:1910.08068
  [hep-ph]} \BibitemShut {NoStop}%
\bibitem [{\citenamefont {Hall}\ \emph {et~al.}(2019)\citenamefont {Hall},
  \citenamefont {Konstandin}, \citenamefont {McGehee},\ and\ \citenamefont
  {Murayama}}]{Hall:2019rld}%
  \BibitemOpen
  \bibfield  {author} {\bibinfo {author} {\bibfnamefont {E.}~\bibnamefont
  {Hall}}, \bibinfo {author} {\bibfnamefont {T.}~\bibnamefont {Konstandin}},
  \bibinfo {author} {\bibfnamefont {R.}~\bibnamefont {McGehee}}, \ and\
  \bibinfo {author} {\bibfnamefont {H.}~\bibnamefont {Murayama}},\ }\href@noop
  {} {\  (\bibinfo {year} {2019})},\ \Eprint {http://arxiv.org/abs/1911.12342}
  {arXiv:1911.12342 [hep-ph]} \BibitemShut {NoStop}%
\bibitem [{\citenamefont {Mangano}\ \emph {et~al.}(2006)\citenamefont
  {Mangano}, \citenamefont {Miele}, \citenamefont {Pastor}, \citenamefont
  {Pinto}, \citenamefont {Pisanti},\ and\ \citenamefont
  {Serpico}}]{Mangano:2006ar}%
  \BibitemOpen
  \bibfield  {author} {\bibinfo {author} {\bibfnamefont {G.}~\bibnamefont
  {Mangano}}, \bibinfo {author} {\bibfnamefont {G.}~\bibnamefont {Miele}},
  \bibinfo {author} {\bibfnamefont {S.}~\bibnamefont {Pastor}}, \bibinfo
  {author} {\bibfnamefont {T.}~\bibnamefont {Pinto}}, \bibinfo {author}
  {\bibfnamefont {O.}~\bibnamefont {Pisanti}}, \ and\ \bibinfo {author}
  {\bibfnamefont {P.~D.}\ \bibnamefont {Serpico}},\ }\href {\doibase
  10.1016/j.nuclphysb.2006.09.002} {\bibfield  {journal} {\bibinfo  {journal}
  {Nucl. Phys. B}\ }\textbf {\bibinfo {volume} {756}},\ \bibinfo {pages} {100}
  (\bibinfo {year} {2006})},\ \Eprint {http://arxiv.org/abs/hep-ph/0607267}
  {arXiv:hep-ph/0607267} \BibitemShut {NoStop}%
\bibitem [{\citenamefont {Fradette}\ \emph {et~al.}(2014)\citenamefont
  {Fradette}, \citenamefont {Pospelov}, \citenamefont {Pradler},\ and\
  \citenamefont {Ritz}}]{Fradette:2014sza}%
  \BibitemOpen
  \bibfield  {author} {\bibinfo {author} {\bibfnamefont {A.}~\bibnamefont
  {Fradette}}, \bibinfo {author} {\bibfnamefont {M.}~\bibnamefont {Pospelov}},
  \bibinfo {author} {\bibfnamefont {J.}~\bibnamefont {Pradler}}, \ and\
  \bibinfo {author} {\bibfnamefont {A.}~\bibnamefont {Ritz}},\ }\href {\doibase
  10.1103/PhysRevD.90.035022} {\bibfield  {journal} {\bibinfo  {journal} {Phys.
  Rev. D}\ }\textbf {\bibinfo {volume} {90}},\ \bibinfo {pages} {035022}
  (\bibinfo {year} {2014})},\ \Eprint {http://arxiv.org/abs/1407.0993}
  {arXiv:1407.0993 [hep-ph]} \BibitemShut {NoStop}%
\bibitem [{\citenamefont {Alexander}\ \emph {et~al.}(2016)\citenamefont
  {Alexander} \emph {et~al.}}]{Alexander:2016aln}%
  \BibitemOpen
  \bibfield  {author} {\bibinfo {author} {\bibfnamefont {J.}~\bibnamefont
  {Alexander}} \emph {et~al.}\ }(\bibinfo {year} {2016})\ \Eprint
  {http://arxiv.org/abs/1608.08632} {arXiv:1608.08632 [hep-ph]} \BibitemShut
  {NoStop}%
\bibitem [{\citenamefont {Chang}\ \emph {et~al.}(2017)\citenamefont {Chang},
  \citenamefont {Essig},\ and\ \citenamefont {McDermott}}]{Chang:2016ntp}%
  \BibitemOpen
  \bibfield  {author} {\bibinfo {author} {\bibfnamefont {J.~H.}\ \bibnamefont
  {Chang}}, \bibinfo {author} {\bibfnamefont {R.}~\bibnamefont {Essig}}, \ and\
  \bibinfo {author} {\bibfnamefont {S.~D.}\ \bibnamefont {McDermott}},\ }\href
  {\doibase 10.1007/JHEP01(2017)107} {\bibfield  {journal} {\bibinfo  {journal}
  {JHEP}\ }\textbf {\bibinfo {volume} {01}},\ \bibinfo {pages} {107} (\bibinfo
  {year} {2017})},\ \Eprint {http://arxiv.org/abs/1611.03864} {arXiv:1611.03864
  [hep-ph]} \BibitemShut {NoStop}%
\bibitem [{\citenamefont {Hardy}\ and\ \citenamefont
  {Lasenby}(2017)}]{Hardy:2016kme}%
  \BibitemOpen
  \bibfield  {author} {\bibinfo {author} {\bibfnamefont {E.}~\bibnamefont
  {Hardy}}\ and\ \bibinfo {author} {\bibfnamefont {R.}~\bibnamefont
  {Lasenby}},\ }\href {\doibase 10.1007/JHEP02(2017)033} {\bibfield  {journal}
  {\bibinfo  {journal} {JHEP}\ }\textbf {\bibinfo {volume} {02}},\ \bibinfo
  {pages} {033} (\bibinfo {year} {2017})},\ \Eprint
  {http://arxiv.org/abs/1611.05852} {arXiv:1611.05852 [hep-ph]} \BibitemShut
  {NoStop}%
\bibitem [{\citenamefont {Pospelov}\ and\ \citenamefont
  {Tsai}(2018)}]{Pospelov:2017kep}%
  \BibitemOpen
  \bibfield  {author} {\bibinfo {author} {\bibfnamefont {M.}~\bibnamefont
  {Pospelov}}\ and\ \bibinfo {author} {\bibfnamefont {Y.-D.}\ \bibnamefont
  {Tsai}},\ }\href {\doibase 10.1016/j.physletb.2018.08.053} {\bibfield
  {journal} {\bibinfo  {journal} {Phys. Lett. B}\ }\textbf {\bibinfo {volume}
  {785}},\ \bibinfo {pages} {288} (\bibinfo {year} {2018})},\ \Eprint
  {http://arxiv.org/abs/1706.00424} {arXiv:1706.00424 [hep-ph]} \BibitemShut
  {NoStop}%
\bibitem [{\citenamefont {Banerjee}\ \emph {et~al.}(2018)\citenamefont
  {Banerjee} \emph {et~al.}}]{Banerjee:2018vgk}%
  \BibitemOpen
  \bibfield  {author} {\bibinfo {author} {\bibfnamefont {D.}~\bibnamefont
  {Banerjee}} \emph {et~al.} (\bibinfo {collaboration} {NA64}),\ }\href
  {\doibase 10.1103/PhysRevLett.120.231802} {\bibfield  {journal} {\bibinfo
  {journal} {Phys. Rev. Lett.}\ }\textbf {\bibinfo {volume} {120}},\ \bibinfo
  {pages} {231802} (\bibinfo {year} {2018})},\ \Eprint
  {http://arxiv.org/abs/1803.07748} {arXiv:1803.07748 [hep-ex]} \BibitemShut
  {NoStop}%
\bibitem [{\citenamefont {Aaij}\ \emph {et~al.}(2018)\citenamefont {Aaij} \emph
  {et~al.}}]{Aaij:2017rft}%
  \BibitemOpen
  \bibfield  {author} {\bibinfo {author} {\bibfnamefont {R.}~\bibnamefont
  {Aaij}} \emph {et~al.} (\bibinfo {collaboration} {LHCb}),\ }\href {\doibase
  10.1103/PhysRevLett.120.061801} {\bibfield  {journal} {\bibinfo  {journal}
  {Phys. Rev. Lett.}\ }\textbf {\bibinfo {volume} {120}},\ \bibinfo {pages}
  {061801} (\bibinfo {year} {2018})},\ \Eprint
  {http://arxiv.org/abs/1710.02867} {arXiv:1710.02867 [hep-ex]} \BibitemShut
  {NoStop}%
\bibitem [{\citenamefont {Aaij}\ \emph {et~al.}(2020)\citenamefont {Aaij} \emph
  {et~al.}}]{Aaij:2019bvg}%
  \BibitemOpen
  \bibfield  {author} {\bibinfo {author} {\bibfnamefont {R.}~\bibnamefont
  {Aaij}} \emph {et~al.} (\bibinfo {collaboration} {LHCb}),\ }\href {\doibase
  10.1103/PhysRevLett.124.041801} {\bibfield  {journal} {\bibinfo  {journal}
  {Phys. Rev. Lett.}\ }\textbf {\bibinfo {volume} {124}},\ \bibinfo {pages}
  {041801} (\bibinfo {year} {2020})},\ \Eprint
  {http://arxiv.org/abs/1910.06926} {arXiv:1910.06926 [hep-ex]} \BibitemShut
  {NoStop}%
\bibitem [{\citenamefont {Parker}\ \emph {et~al.}(2018)\citenamefont {Parker},
  \citenamefont {Yu}, \citenamefont {Zhong}, \citenamefont {Estey},\ and\
  \citenamefont {Müller}}]{Parker:2018vye}%
  \BibitemOpen
  \bibfield  {author} {\bibinfo {author} {\bibfnamefont {R.~H.}\ \bibnamefont
  {Parker}}, \bibinfo {author} {\bibfnamefont {C.}~\bibnamefont {Yu}}, \bibinfo
  {author} {\bibfnamefont {W.}~\bibnamefont {Zhong}}, \bibinfo {author}
  {\bibfnamefont {B.}~\bibnamefont {Estey}}, \ and\ \bibinfo {author}
  {\bibfnamefont {H.}~\bibnamefont {Müller}},\ }\href {\doibase
  10.1126/science.aap7706} {\bibfield  {journal} {\bibinfo  {journal}
  {Science}\ }\textbf {\bibinfo {volume} {360}},\ \bibinfo {pages} {191}
  (\bibinfo {year} {2018})},\ \Eprint {http://arxiv.org/abs/1812.04130}
  {arXiv:1812.04130 [physics.atom-ph]} \BibitemShut {NoStop}%
\bibitem [{\citenamefont {Tsai}\ \emph {et~al.}(2019)\citenamefont {Tsai},
  \citenamefont {deNiverville},\ and\ \citenamefont {Liu}}]{Tsai:2019mtm}%
  \BibitemOpen
  \bibfield  {author} {\bibinfo {author} {\bibfnamefont {Y.-D.}\ \bibnamefont
  {Tsai}}, \bibinfo {author} {\bibfnamefont {P.}~\bibnamefont {deNiverville}},
  \ and\ \bibinfo {author} {\bibfnamefont {M.~X.}\ \bibnamefont {Liu}},\
  }\href@noop {} {\  (\bibinfo {year} {2019})},\ \Eprint
  {http://arxiv.org/abs/1908.07525} {arXiv:1908.07525 [hep-ph]} \BibitemShut
  {NoStop}%
\bibitem [{\citenamefont {Celentano}(2014)}]{Celentano:2014wya}%
  \BibitemOpen
  \bibfield  {author} {\bibinfo {author} {\bibfnamefont {A.}~\bibnamefont
  {Celentano}} (\bibinfo {collaboration} {HPS}),\ }\href {\doibase
  10.1088/1742-6596/556/1/012064} {\bibfield  {journal} {\bibinfo  {journal}
  {J. Phys. Conf. Ser.}\ }\textbf {\bibinfo {volume} {556}},\ \bibinfo {pages}
  {012064} (\bibinfo {year} {2014})},\ \Eprint
  {http://arxiv.org/abs/1505.02025} {arXiv:1505.02025 [physics.ins-det]}
  \BibitemShut {NoStop}%
\bibitem [{\citenamefont {Ilten}\ \emph {et~al.}(2015)\citenamefont {Ilten},
  \citenamefont {Thaler}, \citenamefont {Williams},\ and\ \citenamefont
  {Xue}}]{Ilten:2015hya}%
  \BibitemOpen
  \bibfield  {author} {\bibinfo {author} {\bibfnamefont {P.}~\bibnamefont
  {Ilten}}, \bibinfo {author} {\bibfnamefont {J.}~\bibnamefont {Thaler}},
  \bibinfo {author} {\bibfnamefont {M.}~\bibnamefont {Williams}}, \ and\
  \bibinfo {author} {\bibfnamefont {W.}~\bibnamefont {Xue}},\ }\href {\doibase
  10.1103/PhysRevD.92.115017} {\bibfield  {journal} {\bibinfo  {journal} {Phys.
  Rev. D}\ }\textbf {\bibinfo {volume} {92}},\ \bibinfo {pages} {115017}
  (\bibinfo {year} {2015})},\ \Eprint {http://arxiv.org/abs/1509.06765}
  {arXiv:1509.06765 [hep-ph]} \BibitemShut {NoStop}%
\bibitem [{\citenamefont {Alekhin}\ \emph {et~al.}(2016)\citenamefont {Alekhin}
  \emph {et~al.}}]{Alekhin:2015byh}%
  \BibitemOpen
  \bibfield  {author} {\bibinfo {author} {\bibfnamefont {S.}~\bibnamefont
  {Alekhin}} \emph {et~al.},\ }\href {\doibase 10.1088/0034-4885/79/12/124201}
  {\bibfield  {journal} {\bibinfo  {journal} {Rept. Prog. Phys.}\ }\textbf
  {\bibinfo {volume} {79}},\ \bibinfo {pages} {124201} (\bibinfo {year}
  {2016})},\ \Eprint {http://arxiv.org/abs/1504.04855} {arXiv:1504.04855
  [hep-ph]} \BibitemShut {NoStop}%
\bibitem [{\citenamefont {Ilten}\ \emph {et~al.}(2016)\citenamefont {Ilten},
  \citenamefont {Soreq}, \citenamefont {Thaler}, \citenamefont {Williams},\
  and\ \citenamefont {Xue}}]{Ilten:2016tkc}%
  \BibitemOpen
  \bibfield  {author} {\bibinfo {author} {\bibfnamefont {P.}~\bibnamefont
  {Ilten}}, \bibinfo {author} {\bibfnamefont {Y.}~\bibnamefont {Soreq}},
  \bibinfo {author} {\bibfnamefont {J.}~\bibnamefont {Thaler}}, \bibinfo
  {author} {\bibfnamefont {M.}~\bibnamefont {Williams}}, \ and\ \bibinfo
  {author} {\bibfnamefont {W.}~\bibnamefont {Xue}},\ }\href {\doibase
  10.1103/PhysRevLett.116.251803} {\bibfield  {journal} {\bibinfo  {journal}
  {Phys. Rev. Lett.}\ }\textbf {\bibinfo {volume} {116}},\ \bibinfo {pages}
  {251803} (\bibinfo {year} {2016})},\ \Eprint
  {http://arxiv.org/abs/1603.08926} {arXiv:1603.08926 [hep-ph]} \BibitemShut
  {NoStop}%
\bibitem [{\citenamefont {Caldwell}\ \emph {et~al.}(2018)\citenamefont
  {Caldwell} \emph {et~al.}}]{Caldwell:2018atq}%
  \BibitemOpen
  \bibfield  {author} {\bibinfo {author} {\bibfnamefont {A.}~\bibnamefont
  {Caldwell}} \emph {et~al.},\ }\href@noop {} {\  (\bibinfo {year} {2018})},\
  \Eprint {http://arxiv.org/abs/1812.11164} {arXiv:1812.11164 [physics.acc-ph]}
  \BibitemShut {NoStop}%
\bibitem [{\citenamefont {Altmannshofer}\ \emph {et~al.}(2019)\citenamefont
  {Altmannshofer} \emph {et~al.}}]{Kou:2018nap}%
  \BibitemOpen
  \bibfield  {author} {\bibinfo {author} {\bibfnamefont {W.}~\bibnamefont
  {Altmannshofer}} \emph {et~al.} (\bibinfo {collaboration} {Belle-II}),\
  }\href {\doibase 10.1093/ptep/ptz106} {\bibfield  {journal} {\bibinfo
  {journal} {PTEP}\ }\textbf {\bibinfo {volume} {2019}},\ \bibinfo {pages}
  {123C01} (\bibinfo {year} {2019})},\ \bibinfo {note} {[Erratum: PTEP 2020,
  029201 (2020)]},\ \Eprint {http://arxiv.org/abs/1808.10567} {arXiv:1808.10567
  [hep-ex]} \BibitemShut {NoStop}%
\bibitem [{\citenamefont {Berlin}\ \emph {et~al.}(2018)\citenamefont {Berlin},
  \citenamefont {Gori}, \citenamefont {Schuster},\ and\ \citenamefont
  {Toro}}]{Berlin:2018pwi}%
  \BibitemOpen
  \bibfield  {author} {\bibinfo {author} {\bibfnamefont {A.}~\bibnamefont
  {Berlin}}, \bibinfo {author} {\bibfnamefont {S.}~\bibnamefont {Gori}},
  \bibinfo {author} {\bibfnamefont {P.}~\bibnamefont {Schuster}}, \ and\
  \bibinfo {author} {\bibfnamefont {N.}~\bibnamefont {Toro}},\ }\href {\doibase
  10.1103/PhysRevD.98.035011} {\bibfield  {journal} {\bibinfo  {journal} {Phys.
  Rev. D}\ }\textbf {\bibinfo {volume} {98}},\ \bibinfo {pages} {035011}
  (\bibinfo {year} {2018})},\ \Eprint {http://arxiv.org/abs/1804.00661}
  {arXiv:1804.00661 [hep-ph]} \BibitemShut {NoStop}%
\bibitem [{\citenamefont {Berlin}\ \emph {et~al.}(2019)\citenamefont {Berlin},
  \citenamefont {Blinov}, \citenamefont {Krnjaic}, \citenamefont {Schuster},\
  and\ \citenamefont {Toro}}]{Berlin:2018bsc}%
  \BibitemOpen
  \bibfield  {author} {\bibinfo {author} {\bibfnamefont {A.}~\bibnamefont
  {Berlin}}, \bibinfo {author} {\bibfnamefont {N.}~\bibnamefont {Blinov}},
  \bibinfo {author} {\bibfnamefont {G.}~\bibnamefont {Krnjaic}}, \bibinfo
  {author} {\bibfnamefont {P.}~\bibnamefont {Schuster}}, \ and\ \bibinfo
  {author} {\bibfnamefont {N.}~\bibnamefont {Toro}},\ }\href {\doibase
  10.1103/PhysRevD.99.075001} {\bibfield  {journal} {\bibinfo  {journal} {Phys.
  Rev. D}\ }\textbf {\bibinfo {volume} {99}},\ \bibinfo {pages} {075001}
  (\bibinfo {year} {2019})},\ \Eprint {http://arxiv.org/abs/1807.01730}
  {arXiv:1807.01730 [hep-ph]} \BibitemShut {NoStop}%
\bibitem [{\citenamefont {Ariga}\ \emph {et~al.}(2019)\citenamefont {Ariga}
  \emph {et~al.}}]{Ariga:2018uku}%
  \BibitemOpen
  \bibfield  {author} {\bibinfo {author} {\bibfnamefont {A.}~\bibnamefont
  {Ariga}} \emph {et~al.} (\bibinfo {collaboration} {FASER}),\ }\href {\doibase
  10.1103/PhysRevD.99.095011} {\bibfield  {journal} {\bibinfo  {journal} {Phys.
  Rev. D}\ }\textbf {\bibinfo {volume} {99}},\ \bibinfo {pages} {095011}
  (\bibinfo {year} {2019})},\ \Eprint {http://arxiv.org/abs/1811.12522}
  {arXiv:1811.12522 [hep-ph]} \BibitemShut {NoStop}%
\bibitem [{\citenamefont {NA62}(2018)}]{NA62:2312430}%
  \BibitemOpen
  \bibfield  {author} {\bibinfo {author} {\bibfnamefont {C.}~\bibnamefont
  {NA62}} (\bibinfo {collaboration} {NA62 Collaboration}),\ }\href
  {http://cds.cern.ch/record/2312430} {\emph {\bibinfo {title} {{2018 NA62
  Status Report to the CERN SPSC}}}},\ \bibinfo {type} {Tech. Rep.}\ \bibinfo
  {number} {CERN-SPSC-2018-010. SPSC-SR-229}\ (\bibinfo  {institution} {CERN},\
  \bibinfo {address} {Geneva},\ \bibinfo {year} {2018})\BibitemShut {NoStop}%
\bibitem [{\citenamefont {Harvey}\ and\ \citenamefont
  {Turner}(1990)}]{Harvey:1990qw}%
  \BibitemOpen
  \bibfield  {author} {\bibinfo {author} {\bibfnamefont {J.~A.}\ \bibnamefont
  {Harvey}}\ and\ \bibinfo {author} {\bibfnamefont {M.~S.}\ \bibnamefont
  {Turner}},\ }\href {\doibase 10.1103/PhysRevD.42.3344} {\bibfield  {journal}
  {\bibinfo  {journal} {Phys. Rev. D}\ }\textbf {\bibinfo {volume} {42}},\
  \bibinfo {pages} {3344} (\bibinfo {year} {1990})}\BibitemShut {NoStop}%
\bibitem [{\citenamefont {D'Onofrio}\ \emph {et~al.}(2012)\citenamefont
  {D'Onofrio}, \citenamefont {Rummukainen},\ and\ \citenamefont
  {Tranberg}}]{DOnofrio:2012phz}%
  \BibitemOpen
  \bibfield  {author} {\bibinfo {author} {\bibfnamefont {M.}~\bibnamefont
  {D'Onofrio}}, \bibinfo {author} {\bibfnamefont {K.}~\bibnamefont
  {Rummukainen}}, \ and\ \bibinfo {author} {\bibfnamefont {A.}~\bibnamefont
  {Tranberg}},\ }\href {\doibase 10.1007/JHEP08(2012)123} {\bibfield  {journal}
  {\bibinfo  {journal} {JHEP}\ }\textbf {\bibinfo {volume} {08}},\ \bibinfo
  {pages} {123} (\bibinfo {year} {2012})},\ \Eprint
  {http://arxiv.org/abs/1207.0685} {arXiv:1207.0685 [hep-ph]} \BibitemShut
  {NoStop}%
\bibitem [{\citenamefont {D'Onofrio}\ \emph {et~al.}(2014)\citenamefont
  {D'Onofrio}, \citenamefont {Rummukainen},\ and\ \citenamefont
  {Tranberg}}]{DOnofrio:2014rug}%
  \BibitemOpen
  \bibfield  {author} {\bibinfo {author} {\bibfnamefont {M.}~\bibnamefont
  {D'Onofrio}}, \bibinfo {author} {\bibfnamefont {K.}~\bibnamefont
  {Rummukainen}}, \ and\ \bibinfo {author} {\bibfnamefont {A.}~\bibnamefont
  {Tranberg}},\ }\href {\doibase 10.1103/PhysRevLett.113.141602} {\bibfield
  {journal} {\bibinfo  {journal} {Phys. Rev. Lett.}\ }\textbf {\bibinfo
  {volume} {113}},\ \bibinfo {pages} {141602} (\bibinfo {year} {2014})},\
  \Eprint {http://arxiv.org/abs/1404.3565} {arXiv:1404.3565 [hep-ph]}
  \BibitemShut {NoStop}%
\bibitem [{\citenamefont {Rubakov}\ and\ \citenamefont
  {Shaposhnikov}(1998)}]{Rubakov:1997sr}%
  \BibitemOpen
  \bibfield  {author} {\bibinfo {author} {\bibfnamefont {V.~A.}\ \bibnamefont
  {Rubakov}}\ and\ \bibinfo {author} {\bibfnamefont {M.~E.}\ \bibnamefont
  {Shaposhnikov}},\ }\href {\doibase 10.1063/1.54699} {\bibfield  {journal}
  {\bibinfo  {journal} {AIP Conf. Proc.}\ }\textbf {\bibinfo {volume} {419}},\
  \bibinfo {pages} {347} (\bibinfo {year} {1998})}\BibitemShut {NoStop}%
\bibitem [{\citenamefont {Crewther}\ \emph {et~al.}(1979)\citenamefont
  {Crewther}, \citenamefont {Di~Vecchia}, \citenamefont {Veneziano},\ and\
  \citenamefont {Witten}}]{Crewther:1979pi}%
  \BibitemOpen
  \bibfield  {author} {\bibinfo {author} {\bibfnamefont {R.~J.}\ \bibnamefont
  {Crewther}}, \bibinfo {author} {\bibfnamefont {P.}~\bibnamefont
  {Di~Vecchia}}, \bibinfo {author} {\bibfnamefont {G.}~\bibnamefont
  {Veneziano}}, \ and\ \bibinfo {author} {\bibfnamefont {E.}~\bibnamefont
  {Witten}},\ }\href {\doibase 10.1016/0370-2693(79)90128-X} {\bibfield
  {journal} {\bibinfo  {journal} {Phys. Lett. B}\ }\textbf {\bibinfo {volume}
  {88}},\ \bibinfo {pages} {123} (\bibinfo {year} {1979})},\ \bibinfo {note}
  {[Erratum: Phys.Lett.B 91, 487 (1980)]}\BibitemShut {NoStop}%
\bibitem [{\citenamefont {Zyla}\ \emph
  {et~al.}(2020{\natexlab{a}})\citenamefont {Zyla} \emph
  {et~al.}}]{ParticleDataGroup:2020ssz}%
  \BibitemOpen
  \bibfield  {author} {\bibinfo {author} {\bibfnamefont {P.~A.}\ \bibnamefont
  {Zyla}} \emph {et~al.} (\bibinfo {collaboration} {Particle Data Group}),\
  }\href {\doibase 10.1093/ptep/ptaa104} {\bibfield  {journal} {\bibinfo
  {journal} {PTEP}\ }\textbf {\bibinfo {volume} {2020}},\ \bibinfo {pages}
  {083C01} (\bibinfo {year} {2020}{\natexlab{a}})}\BibitemShut {NoStop}%
\bibitem [{\citenamefont {Hochberg}\ \emph {et~al.}(2018)\citenamefont
  {Hochberg}, \citenamefont {Kuflik},\ and\ \citenamefont
  {Murayama}}]{Hochberg:2017khi}%
  \BibitemOpen
  \bibfield  {author} {\bibinfo {author} {\bibfnamefont {Y.}~\bibnamefont
  {Hochberg}}, \bibinfo {author} {\bibfnamefont {E.}~\bibnamefont {Kuflik}}, \
  and\ \bibinfo {author} {\bibfnamefont {H.}~\bibnamefont {Murayama}},\ }\href
  {\doibase 10.1103/PhysRevD.97.055030} {\bibfield  {journal} {\bibinfo
  {journal} {Phys. Rev. D}\ }\textbf {\bibinfo {volume} {97}},\ \bibinfo
  {pages} {055030} (\bibinfo {year} {2018})},\ \Eprint
  {http://arxiv.org/abs/1706.05008} {arXiv:1706.05008 [hep-ph]} \BibitemShut
  {NoStop}%
\bibitem [{\citenamefont {Kanemura}\ \emph {et~al.}(2015)\citenamefont
  {Kanemura}, \citenamefont {Moroi},\ and\ \citenamefont
  {Tanabe}}]{Kanemura:2015cxa}%
  \BibitemOpen
  \bibfield  {author} {\bibinfo {author} {\bibfnamefont {S.}~\bibnamefont
  {Kanemura}}, \bibinfo {author} {\bibfnamefont {T.}~\bibnamefont {Moroi}}, \
  and\ \bibinfo {author} {\bibfnamefont {T.}~\bibnamefont {Tanabe}},\ }\href
  {\doibase 10.1016/j.physletb.2015.10.002} {\bibfield  {journal} {\bibinfo
  {journal} {Phys. Lett. B}\ }\textbf {\bibinfo {volume} {751}},\ \bibinfo
  {pages} {25} (\bibinfo {year} {2015})},\ \Eprint
  {http://arxiv.org/abs/1507.02809} {arXiv:1507.02809 [hep-ph]} \BibitemShut
  {NoStop}%
\bibitem [{\citenamefont {Liu}\ \emph {et~al.}(2017)\citenamefont {Liu},
  \citenamefont {Wang},\ and\ \citenamefont {Zhang}}]{Liu:2016zki}%
  \BibitemOpen
  \bibfield  {author} {\bibinfo {author} {\bibfnamefont {Z.}~\bibnamefont
  {Liu}}, \bibinfo {author} {\bibfnamefont {L.-T.}\ \bibnamefont {Wang}}, \
  and\ \bibinfo {author} {\bibfnamefont {H.}~\bibnamefont {Zhang}},\ }\href
  {\doibase 10.1088/1674-1137/41/6/063102} {\bibfield  {journal} {\bibinfo
  {journal} {Chin. Phys. C}\ }\textbf {\bibinfo {volume} {41}},\ \bibinfo
  {pages} {063102} (\bibinfo {year} {2017})},\ \Eprint
  {http://arxiv.org/abs/1612.09284} {arXiv:1612.09284 [hep-ph]} \BibitemShut
  {NoStop}%
\bibitem [{\citenamefont {Zyla}\ \emph
  {et~al.}(2020{\natexlab{b}})\citenamefont {Zyla} \emph
  {et~al.}}]{Zyla:2020zbs}%
  \BibitemOpen
  \bibfield  {author} {\bibinfo {author} {\bibfnamefont {P.}~\bibnamefont
  {Zyla}} \emph {et~al.} (\bibinfo {collaboration} {Particle Data Group}),\
  }\href {\doibase 10.1093/ptep/ptaa104} {\bibfield  {journal} {\bibinfo
  {journal} {PTEP}\ }\textbf {\bibinfo {volume} {2020}},\ \bibinfo {pages}
  {083C01} (\bibinfo {year} {2020}{\natexlab{b}})}\BibitemShut {NoStop}%
\end{thebibliography}%

\end{document}